\documentclass[12pt]{article}
\usepackage{amssymb,amsmath,epsfig}

\renewcommand{\theequation}{\arabic{section}.\arabic{equation}}

\begin{document}

\title{\bf Hot Plasma Waves in Schwarzschild Magnetosphere}

\author{M. Sharif \thanks{msharif@math.pu.edu.pk} and Asma Rafique
\thanks{asmarafique@ymail.com}\\
Department of Mathematics,\\
University of the Punjab, Lahore 54590, Pakistan}
\date{}
\maketitle
\begin{abstract}
In this paper we examine the wave properties of hot plasma living in
Schwarzschild magnetosphere. The 3+1 GRMHD perturbation equations
are formulated for this scenario. These equations are Fourier
analyzed and then solved numerically to obtain the dispersion
relations for non-rotating, rotating non-magnetized and rotating
magnetized plasma. The wave vector is evaluated which is used to
calculate refractive index. These quantities are shown in graphs
which are helpful to discuss the dispersive properties of the medium
near the event horizon.
\end{abstract}

\section{Introduction}

In general relativity (GR), the term black hole evokes the
mysterious gravity. Black holes are among the most remarkable
predictions of GR, which are a reality today. These are usually
found in X-ray binaries and in the centers of galaxies. The
existence of rotating stars indicates their rotational behavior. A
rotating black hole gradually reduces to a Schwarzschild
(non-rotating) black hole by the extraction of its rotating
energy. Plasmas are abundant in nature. More than $99\%$ of all
known matter is in the plasma state. All the stars are made of
plasma, and even the space between the stars is filled with
plasma. Commonly space plasma occurs in a hot state. The strong
gravity of the black hole strips the plasma from the surrounding
star. Thus the plasma gathered around the black hole in the form
of accretion disk. The moving plasma creates a magnetic field. The
region surrounding the black hole admitting the magnetic field is
known as magnetosphere. The theory of general relativistic
magnetohydrodynamics (GRMHD) is probably the most accurate
approach to investigate the dynamics of relativistic, magnetized
plasma.

The Schwarzschild black hole is non-rotating and hence the
magnetospheric plasma falls freely along radial direction only.
Perturbations in the Schwarzschild regime, either geometrical or
physical, have always been of interest by the relativists. Regge and
Wheeler \cite{1}, Zerilli \cite{3} and Price \cite{6} discussed the
gravitational perturbations. Fiziev \cite{7} presented the exact
solution of Regge-Wheeler equations. These equations describe the
axial perturbations of the Schwarzschild metric in linear
approximation. The quasi-static problem of electric field in the
Schwarzschild black hole was solved by Hanni and Ruffini \cite{8}.
Sakai and Kawata \cite{9} developed a special relativistic approach
for a linearized treatment of plasma waves in the Schwarzschild
black hole magnetosphere.

In general relativity, a 3+1 hypersurface split of spacetime is
appropriate to understand the black hole physics. This split was
developed by Arnowitt, Deser and Misner (ADM) \cite{10} to study
the quantization of gravitational field. The formalism has wide
applications in numerical relativity. The 3+1 approach has been
used by many authors \cite{13}-\cite{16} to discuss different
features in GR. Thorne and Macdonald \cite{17}-\cite{19} developed
the electromagnetic theory in black hole regime using this
formalism. Holcomb and Tajima \cite{20}, Holcomb \cite{21} and
Dettmann et al. \cite{22} studied some properties of wave
propagation for the Friedmann universe. Buzzi et al. \cite{23}
investigated relativistic two fluid plasma wave properties in the
vicinity of the Schwarzschild black hole. Ali and Rahman \cite{24}
adopted the technique used by Buzzi et al. \cite{23} to analyze
the transverse electromagnetic waves propagating in a plasma close
to the Schwarzschild-de Sitter black hole. Zhang \cite{25}
formulated the black hole theory for stationary symmetric GRMHD.
He \cite{26} also discussed the behavior of cold plasma
perturbations in the Kerr magnetosphere. Sharif et al.
\cite{27}-\cite{33} discussed properties of plasma waves by using
real and complex wave numbers. The analysis was given both for
cold and isothermal plasmas.

A lot of work has been done using cold and isothermal plasma but
no one has used hot plasma in this context which is the basic
constituent of nature. This is the most general plasma which
reduces to cold and isothermal plasma with some restrictions. We
have considered this plasma around the black hole to check the
possibility of receiving information. This work focusses on the
investigation of hot plasma wave properties in the Schwarzschild
magnetosphere. We shall apply perturbation and Fourier analysis
techniques. The dispersion relations are calculated with the help
of the software \emph{Mathematica} to obtain the wave vector. This
will be used to evaluate the refractive index and its change with
respect to angular frequency. The wave properties will be found
through these quantities.

The outline of the paper is as follows. In Section \textbf{2}, we
shall provide the general line element and restrict it to the
Schwarzschild planar analogue. Section \textbf{3} includes the
plasma assumptions for perturbation and Fourier analysis. Moreover,
the perturbed and Fourier analyzed 3+1 GRMHD equations are specified
for hot plasma. In Sections \textbf{4}, \textbf{5} and \textbf{6},
we restrict these equations to non-rotating, rotating non-magnetized
and rotating magnetized plasmas and discuss the wave properties.
Section \textbf{7} contains summary of the results.

\section{3+1 Split of Spacetime}

In \emph{ADM 3+1 split}, the four dimensional spacetime is
decomposed into a succession of three dimensional spacelike
hypersurfaces with the directions normal to them taken to be
\emph{universal time} direction. The line element of spacetime in
3+1 formalism can be written as \cite{26}
$\setcounter{equation}{0}$
\begin{equation}\label{1}
ds^2=-\alpha^2dt^2+\gamma_{ij}(dx^i+\beta^idt)(dx^j+\beta^jdt),
\end{equation}
where $\alpha$ is the lapse function  which describes the ratio of
fiducial proper time to universal time, i.e.,
$\alpha=\frac{d\tau}{dt}$.  The shift vector components $\beta^i$
indicate the shift of spatial coordinates as one moves from one
hypersurface to next. $\gamma_{ij}$ are components of three
dimensional hypersurface (the absolute space) metric. A natural
observer is associated with this spacetime, called the fiducial
observer (FIDO).

The Schwarzschild black hole is non-rotating, thus in the planar
analogue of Schwarzschild geometry, the shift $\beta$ vanishes and
the above line element becomes \cite{31}
\begin{equation}\label{2}
ds^2=-\alpha^2(z)dt^2+dx^2+dy^2+dz^2.
\end{equation}
Here, the directions $z$, $x$ and $y$ are analogous to
Schwarzschild's $r$, $\phi$ and $\theta$ respectively.

\section{3+1 GRMHD Equations with Relative Assumptions}

For the plasma existing in the general and Scwarzschild planar
analogues (given by Eqs.(\ref{1}) and (\ref{2})), the 3+1 GRMHD
equations are given in \textbf{Appendix A}. We assume that hot
plasma surrounds the Schwarzschild black hole. The specific enthalpy
of the fluid is \cite{26}
$\setcounter{equation}{0}$
\begin{eqnarray}\label{3}
\mu=\frac{\rho+p}{\rho_0},
\end{eqnarray}
where $\rho_0,~\rho$ and $p$ denote the rest-mass density, moving
mass density and pressure respectively. Equation (\ref{3}) indicates
the exchange of heat between the plasma and the magnetic field of
the fluid.

We can modify the 3+1 GRMHD equations given by
Eqs.(\ref{54})-(\ref{58}) for hot plasma living in the Schwarzschild
spacetime as follows
\begin{eqnarray}
\label{4} &&\frac{\partial \textbf{B}}{\partial
t}=\nabla\times(\alpha \textbf{V}\times \textbf{B}),\\
\label{5}&&\nabla.\textbf{B}=0,\\
\label{6} &&\frac{\partial (\rho+p) }{\partial t}+(\rho+p)\gamma^2
\textbf{V}. \frac{\partial \textbf{V}}{\partial t}+
(\rho+p)\gamma^2 V.(\alpha \textbf{V}.\nabla)
\textbf{V}\nonumber\\
&&+(\rho+p) \nabla.(\alpha\textbf{V})=0,\\
\label{7}&&\left\{\left((\rho+p)\gamma^2+\frac{\textbf{B}^2}{4\pi}\right)\delta_{ij}
+(\rho+p)\gamma^4V_iV_j-\frac{1}{4\pi}B_iB_j\right\}
\left(\frac{1}{\alpha}\frac{\partial}{\partial
t}\right.\nonumber\\
&&\left.+\textbf{V}.\nabla\right)V^j+\gamma^2V_i(\textbf{V}.\nabla)(\rho+p)
-\left(\frac{\textbf{B}^2}{4\pi}\delta_{ij}-\frac{1}{4\pi}B_iB_j\right)V^j_{,k}V^k\nonumber\\
&&=-(\rho+p)\gamma^2a_i-p_{,i}+\frac{1}{4\pi}
(\textbf{V}\times\textbf{B})_i\nabla.(\textbf{V}\times\textbf{B})
-\frac{1}{8\pi\alpha^2}(\alpha\textbf{B})^2_{,i}\nonumber\\
&&+\frac{1}{4\pi\alpha}(\alpha B_i)_{,j}B^j-\frac{1}{4\pi\alpha}
[\textbf{B}\times\{\textbf{V}\times(\nabla\times(\alpha\textbf{V}\times\textbf{B}))\}]_i,\\
\label{8}&&\frac{1}{\alpha}\frac {\partial}{\partial
t}\{(\rho+p)\gamma^2-p\}+(\rho+p)\gamma^2\textbf{V}.\textbf{a}+
(\rho+p)\gamma^4\textbf{V}.(\textbf{V}.\nabla)\textbf{V}\nonumber
\end{eqnarray}
\begin{eqnarray}
&&-\frac{1}{2\pi}\textbf{a}.\{(\textbf{V}\times\textbf{B})\times\textbf{B}\}
=\frac{1}{4\pi}\textbf{B}.\{\nabla\times(\textbf{V}\times\textbf{B})\}\nonumber\\
&&-\frac{1}{4\pi}\left\{2\frac{1}{\alpha}(\textbf{V}\times\textbf{B}).\frac
{\partial}{\partial t}(\textbf{V}\times\textbf{B})+
(\textbf{V}\times\textbf{B}).(\textbf{a}\times\textbf{B})\right.\nonumber\\
&&\left.+\textbf{B}.\frac{1}{\alpha}\frac {\partial
\textbf{B}}{\partial t}\right\},
\end{eqnarray}
where $\textbf{B}$ and $\textbf{V}$ are the velocity and magnetic
field of the fluid as measured by the FIDO.

We consider rotating background (plasma is rotating), plasma is not
only moving along radial direction but it rotates as well. Due to
its rotation, wave propagates along axial direction also and thus
propagates in $(x,z)$-plane. It is assumed that the FIDO measured
velocity of fluid and magnetic field lie in $xz$-plane
\begin{eqnarray}\label{9}
\textbf{V}=V(z)\textbf{e}_x+u(z)\textbf{e}_z,\quad
\textbf{B}=B[\lambda(z)\textbf{e}_x+\textbf{e}_z],
\end{eqnarray}
where $B$ is an arbitrary constant. Here $\lambda$, $u$ and $V$ are
related by \cite{27}
\begin{equation}\label{a}
V=\frac{V^F}{\alpha}+\lambda u,
\end{equation}
where $V^F$ is an integration constant. The Lorentz factor
$\gamma=\frac{1}{\sqrt{1-\textbf{V}^2}}$ takes the following form
\begin{equation}\label{b}
\gamma=\frac{1}{\sqrt{1-u^2-V^2}}.
\end{equation}
The plasma flow in the magnetosphere can be characterized by its
density $\rho$, pressure $p$, velocity $\textbf{V}$ and magnetic
field $\textbf{B}$. When the flow is perturbed, these variables will
become
\begin{eqnarray}\label{11}
&&\rho=\rho^0+\delta\rho=\rho^0+\rho\widetilde{\rho},\quad
p=p^0+\delta p=p^0+p\widetilde{p},\nonumber\\
&&\textbf{V}=\textbf{V}^0+\delta\textbf{V}=\textbf{V}^0+\textbf{v},\quad
\textbf{B}=\textbf{B}^0+\delta\textbf{B}=\textbf{B}^0+B\textbf{b},
\end{eqnarray}
where $\rho^0,~p^0,~\textbf{V}^0$ and $\textbf{B}^0$ are
unperturbed quantities. The linear perturbations in these
quantities are denoted by  $\delta \rho$, $\delta p$,
$\delta\textbf{V}$ and $\delta\textbf{B}$ respectively.

We shall use the following dimensionless notations for the perturbed
quantities
\begin{eqnarray}\label{12}
&&\tilde{\rho}=\tilde{\rho}(t,z),\quad \tilde{p}=\tilde{p}(t,z),\nonumber\\
&&\textbf{v}=\delta\textbf{V}=v_x(t,z)\textbf{e}_x
+v_z(t,z)\textbf{e}_z,\nonumber\\
&&\textbf{b}=\frac{\delta\textbf{B}}{B}=b_x(t,z)\textbf{e}_x
+b_z(t,z)\textbf{e}_z,
\end{eqnarray}
where $\widetilde{\rho},~\widetilde{p}, v_x, v_z, b_x$ and $b_z$ are dimensionless quantities.
For Fourier analysis, we assume the harmonic space and time
dependence of perturbations
\begin{eqnarray}\label{13}
\widetilde{\rho}(t,z)=c_1e^{-\iota(\omega t-kz)},&\quad&
\widetilde{p}(t,z)=c_2e^{-\iota(\omega t-kz)},\nonumber\\
v_z(t,z)=c_3e^{-\iota(\omega t-kz)},&\quad&
v_x(t,z)=c_4e^{-\iota(\omega t-kz)},\nonumber\\
b_z(t,z)=c_5e^{-\iota(\omega t-kz)},&\quad&
b_x(t,z)=c_6e^{-\iota(\omega t-kz)},
\end{eqnarray}
where $k$ is $z$-component of the wave vector $(0,0,k)$ and $\omega$
is the angular frequency. We can use the wave vector to obtain the
refractive index which we shall use to deduce the plasma wave
properties. We define the wave vector and the refractive index as
follows:
\begin{itemize}
\item\textbf{Wave Vector}: A wave vector is a vector which points
in the direction of propagation of wave. Its magnitude gives the
wave number.
\item\textbf{Refractive Index}: It is the ratio of the speed of light in
vacuum to the speed of light in the other medium \cite{36}. If the
refractive index is greater than one and its change with respect
to angular frequency is positive, the dispersion is said to be
normal, otherwise anomalous.
\end{itemize}

When we introduce linear perturbations from Eq.(\ref{11}), the
perfect GRMHD Eqs.(\ref{4})-(\ref{8}) turn out to be
\begin{eqnarray}
\label{14}&&\frac{\partial(\delta\textbf{B})}{\partial
t}=\nabla\times(\alpha\textbf{v}\times\textbf{B})
+\nabla\times(\alpha\textbf{V}\times\delta\textbf{B}),\\
\label{15}&&\nabla.(\delta\textbf{B})=0,\\
\label{16}&&\frac{1}{\alpha}\frac{\partial(\delta\rho+\delta
p)}{\partial
t}+(\rho+p)\gamma^2\textbf{V}.(\frac{1}{\alpha}\frac{\partial}{\partial
t}+\textbf{V}.\nabla)\textbf{v}\nonumber+(\rho+p)(\nabla.\textbf{v})\nonumber\\
&&=-2(\rho+p)\gamma^2(\textbf{V}.\textbf{v})(\textbf{V}.\nabla)\ln\gamma-
(\rho+p)\gamma^2(\textbf{V}.\nabla\textbf{V}).\textbf{v}\nonumber\\
&&+(\rho+p)(\textbf{v}.\nabla\ln u),\\
\label{17}&&\left\{\left((\rho+p)\gamma^2
+\frac{\textbf{B}^2}{4\pi}\right)\delta_{ij}+(\rho+p)\gamma^4V_iV_j
-\frac{1}{4\pi}B_iB_j\right\}\frac{1}{\alpha}\frac{\partial
v^j}{\partial t}\nonumber\\
&&+\frac{1}{4\pi}[\textbf{B}\times\{\textbf{V}
\times\frac{1}{\alpha}\frac{\partial(\delta\textbf{B})}{\partial
t}\}]_i+(\rho+p)\gamma^2v_{i,j}V^j+(\rho+p)\nonumber\\
&&\times\gamma^4V_iv_{j,k}V^jV^k
+\gamma^2V_i(\textbf{V}.\nabla)(\delta \rho+\delta
p)+\gamma^2V_i(\textbf{v}.\nabla)(\rho+p)\nonumber\\
&&+\gamma^2v_i(\textbf{V}.\nabla)(\rho+p)
+\gamma^4(2\textbf{V}.\textbf{v})V_i(\textbf{V}.\nabla)(\rho+p)
-\frac{1}{4\pi\alpha}\{(\alpha\delta
B_i)_{,j}\nonumber\\
&&-(\alpha\delta B_j)_{,i}\}B^j=-(\delta
p)_i-\gamma^2\{(\delta\rho+\delta
p)+2(\rho+p)\gamma^2(\textbf{V}.\textbf{v})\}a_i \nonumber
\end{eqnarray}
\begin{eqnarray}
&&+\frac{1}{4\pi\alpha}\{(\alpha B_i)_{,j}-(\alpha B_
j)_{,i}\}\delta
B^j-(\rho+p)\gamma^4(v_iV^j+v^jV_i)V_{k,j}V^k\nonumber\\
&&-\gamma^2\{(\delta\rho+\delta
p)V^j+2(\rho+p)\gamma^2(\textbf{V}.\textbf{v})V^j+(\rho+p)v^j\}V_{i,j}\nonumber\\
&&-\gamma^4V_i\{(\delta\rho+\delta
p)V^j+4(\rho+p)\gamma^2(\textbf{V}.\textbf{v})V^j \nonumber\\
&&+(\rho+p)v^j\}V_{j,k}V^k,\\
\label{18}&&\gamma^2\frac{1}{\alpha}\frac{\partial(\delta\rho+\delta
p)}{\partial
t}+\frac{2}{\alpha}(\rho+p)\gamma^4(\textbf{V}.\frac{\partial\textbf{v}}{\partial
t })-\frac{1}{\alpha}\frac{\partial(\delta p)}{\partial
t}+(\delta\rho+\delta p)\gamma^2\nonumber\\
&&\times[(\textbf{V}.\textbf{a})
+\gamma^2\textbf{V}.(\textbf{V}.\nabla)\textbf{V}]
+(\rho+p)\gamma^2[\textbf{v}.\textbf{a}
+2\gamma^2(\textbf{V}.\textbf{v})(\textbf{V}.\textbf{a})
\nonumber\\
&&+4\gamma^4(\textbf{V}.\textbf{v})\textbf{V}.(\textbf{V}.\nabla)\textbf{V}
+\gamma^2(\textbf{v}.(\textbf{V}.\nabla)\textbf{V}
+\textbf{V}.(\textbf{v}.\nabla)\textbf{V}
+\textbf{V}.(\textbf{V}.\nabla)\textbf{v})]\nonumber\\
&&=\frac{1}{4\pi}[\delta\textbf{B}.\nabla\times(\textbf{V}\times\textbf{B})
+\textbf{B}.\nabla\times(\textbf{v}\times\textbf{B})
+\textbf{B}.\nabla\times(\textbf{V}\times\delta\textbf{B})\nonumber\\
&&-\frac{2}{\alpha}(\textbf{v}\times\textbf{B}
+\textbf{V}\times\delta\textbf{B}).\frac{\partial}{\partial
t}(\textbf{V}\times\textbf{B})
-\frac{2}{\alpha}(\textbf{V}\times\textbf{B}).\left(\frac{\partial
\textbf{v}}{\partial
t}\times\textbf{B}\right.\nonumber\\
&&\left.+\textbf{V}\times\frac{\partial\delta\textbf{B}}{\partial
t}\right)-(\textbf{v}\times\textbf{B}
+\textbf{V}\times\delta\textbf{B}).(\textbf{a}\times\textbf{B})
-(\textbf{V}\times\textbf{B}).(\textbf{a}\times\delta\textbf{B})\nonumber\\
&&-\frac{\delta\textbf{B}}{\alpha}.\frac{\partial\textbf{B}}{\partial
t}-\frac{\textbf{B}}{\alpha}.\frac{\partial(\delta\textbf{B})}{\partial
t}]+\frac{1}{2\pi}\textbf{a}.[(\textbf{v}\times\textbf{B})\times\textbf{B}
+(\textbf{V}\times\delta\textbf{B})\times\textbf{B}\nonumber\\
&&+(\textbf{V}\times\textbf{B})\times\delta\textbf{B}].
\end{eqnarray}

When we use assumptions given by Eq.(\ref{12}), the component form
of Eqs.(\ref{14})-(\ref{18}) can be written as
\begin{eqnarray}
\label{19}&&\frac{1}{\alpha}\frac{\partial b_x}{\partial
t}+ub_{x,z}=(v_x-\lambda v_z+Vb_z-ub_x)\nabla
\ln\alpha\nonumber\\
&&+(v_{x,z}-\lambda
v_{z,z}-\lambda'v_z+V'b_z+Vb_{z,z}-u'b_x),\\
\label{20}&&\frac{1}{\alpha}\frac{\partial b_z}{\partial t}+ub_{z,z}=0,\\
\label{21}&&b_{z,z}=0,\\
\label{22}&&\rho\frac{1}{\alpha}\frac{\partial\tilde{\rho}}{\partial
t} +p\frac{1}{\alpha}\frac{\partial\tilde{p}}{\partial
t}+(\rho+p)\gamma^2V(\frac{1}{\alpha}\frac{\partial{v_x}}{\partial
t}+uv_{x,z})+(\rho+p)\gamma^2u\nonumber\\
&&\times\frac{1}{\alpha}\frac{\partial{v_z}}{\partial
t}+(\rho+p)(1+\gamma^2u^2)v_{z,z}=-\gamma^2u(\rho+p)[(1+2\gamma^2V^2)V'\nonumber\\
&&+2\gamma^2uVu']v_x+(\rho+p)[(1-2\gamma^2u^2)(1+\gamma^2u^2)\frac{u'}{u}\nonumber\\
&&-2\gamma^4u^2VV']v_z,
\end{eqnarray}
\begin{eqnarray}
\label{23}&&\left\{(\rho+p)\gamma^2(1+\gamma^2V^2)
+\frac{B^2}{4\pi}\right\}\frac{1}{\alpha}\frac{\partial
v_x}{\partial t}+\left\{(\rho+p)\gamma^4uV-\frac{\lambda B
^2}{4\pi}\right\}\nonumber\\
&&\times\frac{1}{\alpha}\frac{\partial v_z}{\partial
t}+\left\{(\rho+p)\gamma^2(1+\gamma^2V^2)
-\frac{B^2}{4\pi}\right\}uv_{x,z}+\left\{(\rho+p)\gamma^4uV\right.\nonumber\\
&&\left.+\frac{\lambda B^2}{4\pi}\right\}uv_{z,z}
-\frac{B^2}{4\pi}(1-u^2)b_{x,z}-\frac{B^2}{4\pi\alpha}\left\{\alpha'(1-u^2)-\alpha
uu'\right\}b_x\nonumber\\
&&+\gamma^2u(\rho\tilde{\rho}+p\tilde{p})\left\{(1+\gamma^2V^2)V'+\gamma^2uVu'\right\}
+\gamma^2uV(\rho'\tilde{\rho}+\rho\tilde{\rho}'\nonumber\\
&&+p'\tilde{p}+p\tilde{p}')+[(\rho+p)\gamma^4u
\left\{(1+4\gamma^2V^2)uu'+4VV'(1+\gamma^2V^2)\right\}\nonumber\\
&&+\frac{B^2u\alpha'}{4\pi\alpha}
+\gamma^2u(1+2\gamma^2V^2)(\rho'+p')]v_x
+[(\rho+p)\gamma^2\left\{(1+2\gamma^2u^2)\right.\nonumber\\
&&\left.(1+2\gamma^2V^2)V'-\gamma^2V^2V'
+2\gamma^2(1+2\gamma^2u^2)uVu'\right\}+\frac{B^2u}
{4\pi\alpha}(\lambda\alpha)'\nonumber\\
&&+\gamma^2V(1+2\gamma^2u^2)(\rho'+p')]v_z=0,\\
\label{24}&&\left\{(\rho+p)\gamma^2(1+\gamma^2u^2)
+\frac{\lambda^2B^2}{4\pi}\right\}\frac{1}{\alpha}\frac{\partial
v_z}{\partial t}+\left\{(\rho+p)\gamma^4uV -\frac{\lambda B
^2}{4\pi}\right\}\nonumber\\
&&\times\frac{1}{\alpha}\frac{\partial v_x}{\partial t}
+\left\{(\rho+p)\gamma^2(1+\gamma^2u^2)-\frac{\lambda^2B^2}{4\pi}\right\}
uv_{z,z}+\left\{(\rho+p)\gamma^4uV \right.\nonumber\\
&&\left.+\frac{\lambda B^2}{4\pi}\right\}uv_{x,z}+\frac{\lambda
B^2}{4\pi}(1-u^2)b_{x,z}+\frac{B^2}{4\pi\alpha}\left\{(\alpha\lambda)'
+\alpha'\lambda-u\lambda(u\alpha'\right.\nonumber\\
&&\left.+u'\alpha)\right\}b_x+(\rho\tilde{\rho}+p\tilde{p})\gamma^2\left\{a_z
+uu'(1+\gamma^2u^2)+\gamma^2u^2VV'\right\}\nonumber\\
&&+(1+\gamma^2u^2)(p'\tilde{p}+p\tilde{p}')
+\gamma^2u^2(\rho'\tilde{\rho}+\rho\tilde{\rho}')+[(\rho+p)\gamma^4\nonumber\\
&&\times\{u^2V'(1+4\gamma^2V^2)+2V(a_z+uu'(1+2\gamma^2u^2))\}+\frac{\lambda
B^2u\alpha'}{4\pi\alpha}\nonumber\\
&&+2\gamma^4u^2V(\rho'+p')]v_x+[(\rho+p)\gamma^2
\left\{u'(1+\gamma^2u^2)(1+4\gamma^2u^2)\right.\nonumber\\
&&\left.+2u\gamma^2(a_z+(1+2\gamma^2u^2)VV')\right\}
-\frac{\lambda B^2u}{4\pi\alpha}(\alpha\lambda)'
+2\gamma^2u(1\nonumber\\
&&+\gamma^2u^2)(\rho'+p')]v_z=0,\\
\label{25}&&\gamma^2\rho\frac{1}{\alpha}\frac{\partial\tilde{\rho}}{\partial
t}+(\gamma^2-1)p\frac{1}{\alpha}\frac{\partial\tilde{p}}{\partial
t}+\frac{2}{\alpha}\left\{(\rho+p)\gamma^4V-\frac{B^2}{4\pi}(\lambda
u -V)\right\}\nonumber\\
&&\times\frac{\partial v_x}{\partial
t}+\frac{2}{\alpha}\left\{(\rho+p)\gamma^4u+\frac{\lambda
B^2}{4\pi}(\lambda u-V)\right\}\frac{\partial v_z}{\partial
t}+\frac{ B^2}{4\pi\alpha}\left\{\lambda\right.\nonumber\\
&&\left.+2u(\lambda u-V)\right\}\frac{\partial b_x}{\partial t}
+\frac{ B^2}{4\pi\alpha}\left\{1-2V(\lambda
u-V)\right\}\frac{\partial b_z}{\partial t}
+(\rho\tilde{\rho}+p\tilde{p})\nonumber\\
&&\times\gamma^2u\left\{a_z+\gamma^2(VV'+uu')\right\}
+\left\{(\rho+p)\gamma^4uV-\frac{\lambda
B^2}{4\pi}\right\}v_{x,z}\nonumber
\end{eqnarray}
\begin{eqnarray}
&&+\left\{(\rho+p)\gamma^4u^2+\frac{\lambda^2
B^2}{4\pi}\right\}v_{z,z}
+\frac{\lambda B^2u}{4\pi}b_{x,z}
-\frac{\lambda B^2V}{4\pi}b_{z,z}\nonumber\\
&&+[(\rho+p)\gamma^2\left\{(a_z+2\gamma^2uu')(1+2\gamma^2u^2)
+\gamma^2VV'(1+4\gamma^2u^2)\right\}\nonumber\\
&&+\frac{\lambda B^2}{4\pi}(3\lambda
a_z+\lambda')]v_z+[(\rho+p)\gamma^4u
\left\{V'(1+4\gamma^2V^2)+2V\right.\nonumber\\
&&\left.(a_z+2\gamma^2uu')\right\}
-\frac{3 B^2}{4\pi}a_z\lambda]v_x+\frac{B^2}{4\pi}[3a_z(2\lambda
u-V)+2u'\lambda\nonumber\\
&&+u\lambda'-V']b_x-\frac{\lambda B^2}{4\pi}(3Va_z+V')b_z=0.
\end{eqnarray}
Using time dependence harmonic space of the perturbed variables from
Eq.(\ref{13}), we get the Fourier analyzed form of
Eqs.(\ref{19})-(\ref{25})
\begin{eqnarray}
\label{26}&&c_4(\alpha'+\iota
k\alpha)-c_3\{(\alpha\lambda)'+\iota k\alpha\lambda\}+c_5(\alpha
V)'-c_6\{(\alpha
u)'-\iota\omega\nonumber\\
&&+\iota ku\alpha\}=0,\\
\label{27}&&c_5(\frac{-\iota\omega}{\alpha}+\iota k\alpha)=0,\\
\label{28}&&c_5\iota k=0,\\
\label{29}&&c_1(\frac{-\iota\omega}{\alpha}\rho)+c_2(\frac{-\iota\omega}{\alpha}p)
+c_3(\rho+p)[\frac{-\iota\omega}{\alpha}\gamma^2u+(1+\gamma^2u^2)\iota k\nonumber\\
&&-(1-2\gamma^2u^2)(1+\gamma^2u^2)\frac{u'}{u}+2\gamma^4u^2VV']
+c_4(\rho+p)\gamma^2[(\frac{-\iota\omega}{\alpha}\nonumber\\
&&+\iota ku)V+u(1+2\gamma^2V^2)V'+2\gamma^2u^2Vu']=0,\\
\label{30}&&c_1[\rho\gamma^2u\{(1+\gamma^2V^2)V'+\gamma^2Vuu'\}+\gamma^2Vu(\rho'+\iota
k\rho)]\nonumber\\
&&+c_2[p\gamma^2u\{(1+\gamma^2V^2)V'+\gamma^2Vuu'\}+\gamma^2Vu(p'+\iota kp)]\nonumber\\
&&+c_3[(\rho+p)\gamma^2
\{(1+2\gamma^2u^2)(1+2\gamma^2V^2)V'+(\frac{-\iota\omega}{\alpha}+\iota
ku)\gamma^2Vu\nonumber\\
&&-\gamma^2V^2V'+2\gamma^2(1+2\gamma^2u^2)uVu'\}+\gamma^2V(1+2\gamma^2u^2)(\rho'+p')\nonumber\\
&&+\frac{B^2u}{4\pi\alpha}(\lambda\alpha)'+\frac{\lambda
B^2}{4\pi}(\frac{\iota\omega}{\alpha}+\iota
ku)]+c_4[(\rho+p)\gamma^4u\{(1+4\gamma^2V^2)\nonumber\\ &&\times
uu'+4VV'(1+\gamma^2V^2)\}
+(\rho+p)\gamma^2(1+\gamma^2V^2)(\frac{-\iota\omega}{\alpha}+\iota
ku)\nonumber\\
&&+\gamma^2u(1+2\gamma^2V^2)(\rho'+p')+\frac{B^2u\alpha'}{4\pi\alpha}
-\frac{B^2}{4\pi}(\frac{\iota\omega}{\alpha}+\iota
ku)]\nonumber\\
&&+c_6\frac{B^2}{4\pi\alpha}[\alpha uu'-\alpha'(1-u^2)-(1-u^2)\iota k\alpha]=0,\\
\label{31}&&c_1[\rho\gamma^2\{a_z+(1+\gamma^2u^2)uu'+\gamma^2u^2VV'\}+\gamma^2u^2(\rho'+\iota
k\rho)]\nonumber
\end{eqnarray}
\begin{eqnarray}
&&+c_2[p\gamma^2\{a_z+(1+\gamma^2u^2)uu'+\gamma^2u^2VV'\}+(1+\gamma^2u^2)\nonumber\\
&&\times(p'+\iota kp)]+c_3[(\rho+p)\gamma^2
\{(1+\gamma^2u^2)(\frac{-\iota\omega}{\alpha}+\iota ku)\nonumber\\
&&+u'(1+\gamma^2u^2)(1+4\gamma^2u^2)+2u\gamma^2(a_z
+(1+2\gamma^2u^2)VV')\}\nonumber\\
&&+2\gamma^2u(1+\gamma^2u^2)(\rho'+p')-\frac{\lambda
B^2u}{4\pi\alpha}(\lambda\alpha)'-\frac{\lambda^2
B^2}{4\pi}(\frac{\iota\omega}{\alpha}+\iota ku)]\nonumber\\
&&+c_4[(\rho+p)\gamma^4\{(\frac{-\iota\omega}{\alpha}+\iota
ku)uV+u^2V'(1+4\gamma^2V^2)+2V(a_z\nonumber\\
&&+(1+2\gamma^2u^2)uu')\}+2\gamma^4u^2V(\rho'+p')+\frac{\lambda
B^2}{4\pi}(\frac{\iota\omega}{\alpha}+\iota ku)\nonumber\\
&&+\frac{\lambda B^2u\alpha'}{4\pi\alpha}]
+c_6[\frac{B^2}{4\pi\alpha}\{(\lambda\alpha)'
+\alpha'\lambda-u\lambda(u\alpha'+u'\alpha)\}\nonumber\\
&&+\frac{\lambda B^2}{4\pi}(1-u^2)\iota k]=0,\\
\label{32}&&c_1\rho\gamma^2[-\frac{\iota\omega}{\alpha}
+u\{a_z+\gamma^2(VV'+uu')\}]
+c_3[-\frac{2\iota\omega}{\alpha}\{(\rho+p)\gamma^4u\nonumber\\
&&+\frac{B^2\lambda}{4\pi}(u\lambda-V)\}+\iota
k\{(\rho+p)\gamma^4u^2+\frac{\lambda^2
B^2}{4\pi}\}+(\rho+p)\gamma^2\{(a_z\nonumber\\
&&+2\gamma^2uu')
(1+2\gamma^2u^2)+\gamma^2VV'(1+4\gamma^2u^2)\}+(3\lambda
a_z+\lambda')\frac{\lambda B^2}{4\pi}]\nonumber\\
&&+c_4[-\frac{2\iota\omega}{\alpha}\{(\rho+p)\gamma^4V
-\frac{B^2}{4\pi}(u\lambda-V)\}+\iota
k\{(\rho+p)\gamma^4Vu\nonumber\\
&&-\frac{\lambda
B^2}{4\pi}\}+(\rho+p)\gamma^4u\{V'(1+4\gamma^2V^2)+2V(a_z+2\gamma^2uu')\}\nonumber\\
&&-\frac{3B^2}{4\pi}a_z\lambda]+c_6\frac{B^2}{4\pi}[-\frac{\iota\omega}{\alpha}
\{\lambda+2u(u\lambda-V)\}+\iota ku\lambda
+2u'\lambda\nonumber\\
&&+3a_z(2\lambda
u-V)+u\lambda'-V']+c_2p[-\frac{\iota\omega}{\alpha}(\gamma^2-1)
+u\gamma^2\{a_z\nonumber\\
&&+\gamma^2(VV'+uu')\}]=0.
\end{eqnarray}
Equations (\ref{27}) and (\ref{28}) give $c_5=0$ indicating that
there are no perturbations in the $z$-component of magnetic field.

\section{Non-Rotating Plasma Flow}

In non-rotating plasma flow, the magnetospheric perturbations are
only along $z$-axis. The FIDO measured magnetic field and velocity
admit only $z$-component i.e., $V=0=\lambda$ which gives
$v_x=0=b_x$. The Fourier analyzed perturbed GRMHD equations for
non-rotating plasma can be obtained by substituting these
assumptions along with the vanishing respective Fourier constants
$c_4$ and $c_6$ in Eqs.(\ref{26})-(\ref{32})
\begin{eqnarray}{\setcounter{equation}{1}}
\label{36}
&&-\frac{\iota\omega}{\alpha}c_5=0,\\
\label{37}&&\iota kc_5=0,\\
\label{38}&&c_1\{\frac{-\iota\omega}{\alpha}\rho\}
+c_2\{\frac{-\iota\omega}{\alpha}p\}+c_3(\rho+p)\{(1+\gamma^2u^2)\iota
k-\frac{-\iota\omega}{\alpha}\gamma^2u\nonumber\\
&&-(1-2\gamma^2u^2)(1+\gamma^2u^2)\frac{u'}{u}\}=0,\\
\label{39}&&c_1\{\rho\gamma^2(a_z+uu'(1+\gamma^2u^2))+\gamma^2u^2(\rho'+\iota
k\rho)\}\nonumber\\
&&+c_2\{p\gamma^2(a_z +u
u'(1+\gamma^2u^2))+(1+\gamma^2u^2)(p'+\iota kp)\}\nonumber\\
&&+c_3\gamma^2[(\rho+p)\{(1+\gamma^2u^2)(\frac{-\iota\omega}{\alpha}+\iota
ku)+u'(1+\gamma^2u^2)\nonumber\\
&&\times(1+4\gamma^2u^2)+2\gamma^2ua_z\}+2u(1+\gamma^2u^2)(\rho'+p')]=0,\\
\label{40}&&c_1\rho\gamma^2\{\frac{-\iota\omega}{\alpha}+u(a_z+\gamma^2uu')\}
+c_2p\{\frac{-\iota\omega}{\alpha}(\gamma^2-1)\nonumber\\
&&+\gamma^2u(a_z+\gamma^2uu')\}
+c_3(\rho+p)\gamma^2\{\frac{-2\iota\omega}{\alpha}\gamma^2u\nonumber\\
&&+\gamma^2u^2\iota k+(a_z+2\gamma^2uu')(1+2\gamma^2u^2)\}=0.
\end{eqnarray}
We shall use these equations to obtain dispersion relations.

\subsection{Numerical Solutions}

To obtain numerical solutions, we use the following assumptions
\begin{itemize}
\item Time lapse: $\alpha=\tanh(10z)/10,$
\item Specific enthalpy: $\mu=\sqrt{\frac{1-(\tanh(10z)/10)^2}{2}},$
\item Stationary fluid: $\alpha\gamma=1~\Rightarrow\gamma=1/\sqrt{1-u^2}=1/\alpha$
\item For the fluid, freely falling towards the black hole,
the $z$-component of velocity is taken to be
$u=-\sqrt{1-\alpha^2}$.
\item Stiff fluid: $\rho=p=\mu/2$.
\end{itemize}
Using these assumptions, we solve the determinant of the
coefficients of constants of Eqs.(\ref{36})-(\ref{40}) which
results a complex dispersion relation \cite{35}. Comparing the
real and imaginary parts, two dispersion relations are obtained.
The real part gives a relation of the form
\begin{equation}\label{nr1}
A_1(z,\omega)k+A_2(z,\omega)=0
\end{equation}
which is linear and yields only one value of $k$. The imaginary part
gives a dispersion relation quadratic in $k$, i.e., of the type
\begin{equation}\label{nr2}
B_1(z,\omega)k^2+B_2(z,\omega)k+B_3(z,\omega)=0.
\end{equation}
Using the values of $k$, we can calculate the refractive index and
its change with respect to angular frequency which helps us to
study the wave properties.

Figure 1 shows the graphs of $k$ obtained from Eq.(\ref{nr1})
whereas Figures 2 and 3 represent values of $k$ obtained from
Eq.(\ref{nr2}). The graph labels A, B and C denote the graphs of
the wave vector, the refractive index and change in the refractive
index with respect to angular frequency $\omega$ respectively.

Here we are give a brief description of Figure $1$ which helps to
access the results for the other figures.

In Figure 1, the wave vector admits positive values which
indicates that the waves move away from the event horizon. The
refractive index is greater than one throughout the region and
increases in a small region $1.2\leq z\leq1.8,0\leq\omega\leq 10$
with the decrease in $z$. The change in refractive index with
respect to angular frequency is positive at random points which
shows that the waves disperse normally at those points. The
dispersion is anomalous at rest of the points due to negative
values of $\frac{dn}{d\omega}$.\\The information obtained from
Figures 1-3 is given in the following table.
\begin{center}
Table I: Direction and refractive index of waves
\end{center}
\begin{tabular}{|c|c|c|c|c|}
\hline & \textbf{Direction of waves} & \textbf{Refractive index}
($n$)\\ \hline& & $n>1$ and increases in the region \\
\textbf{1} & Move away from the event horizon & $1.2\leq z\leq1.8,0\leq\omega\leq 10$\\
&&with the decrease in $z$  \\
\hline
\textbf{2}& Move towards the event horizon&  Same as above\\
\hline
\textbf{3} & Same as $1$ & Same as above\\
\hline
\end{tabular}
\\
\\
\\
In Figures $1,2$ and $3$, dispersion is normal as well as
anomalous at random points.

\section{Rotating Non-Magnetized Hot Plasma}

This section is devoted to rotating non-magnetized hot plasma
flow. When we consider non-magnetized plasma in rotating
background, i.e., $\textbf{B}=0$, the equations of evolution of
magnetic field (\ref{4}) and (\ref{5}) are satisfied. We
substitute $c_5=0,~B=0,~\lambda=0$ and $c_6=0$ in the Fourier
analyzed perturbed GRMHD Eqs.(\ref{29})-(\ref{32}) and obtain
\begin{eqnarray}{\setcounter{equation}{1}}
\label{46}&&c_1(\frac{-\iota\omega}{\alpha}\rho)+c_2(\frac{-\iota\omega}{\alpha}p)
+c_3(\rho+p)[\frac{-\iota\omega}{\alpha}\gamma^2u+(1+\gamma^2u^2)\iota
k\nonumber\\
&&-(1-2\gamma^2u^2)(1+\gamma^2u^2)\frac{u'}{u}+2\gamma^4u^2VV']
+c_4(\rho+p)\gamma^2[(\frac{-\iota\omega}{\alpha}\nonumber\\
&&+\iota ku)V +u(1+2\gamma^2V^2)V'+2\gamma^2u^2Vu']=0,\\
\label{47}&&c_1[\rho\gamma^2
u\{(1+\gamma^2V^2)V'+\gamma^2Vuu'\}+\gamma^2Vu(\rho'+\iota
k\rho)]+c_2[p\gamma^2u\nonumber\\
&&\times\{(1+\gamma^2V^2)V'+\gamma^2Vuu'\}+\gamma^2Vu(p'+\iota
kp)]+c_3[(\rho+p)\gamma^2\nonumber\\
&&\times\{(\frac{-\iota\omega}{\alpha}+\iota
ku)\gamma^2Vu+(1+2\gamma^2u^2)(1+2\gamma^2V^2)V'-\gamma^2V^2V'\nonumber\\
&&+2\gamma^2(1+2\gamma^2u^2)uVu'\}
+\gamma^2V(1+2\gamma^2u^2)(\rho'+p')]+c_4[(\rho+p)\nonumber\\
&&\{\gamma^2(1+\gamma^2V^2)(\frac{-\iota\omega}{\alpha}+\iota
ku)+\gamma^4u((1+4\gamma^2V^2)uu'+4VV'(1+\nonumber\\
&&\gamma^2V^2))\} +\gamma^2u(1+2\gamma^2V^2)(\rho'+p')]=0,\\
\label{48}&&c_1[\rho\gamma^2\{a_z+(1+\gamma^2u^2)uu'
+\gamma^2u^2VV'\}+\gamma^2u^2(\rho'+\iota
k\rho)]\nonumber\\
&&+c_2[p\gamma^2\{a_z
+(1+\gamma^2u^2)uu'+\gamma^2u^2VV'\}+(p'+\iota
kp)\times\nonumber\\
&&(1+\gamma^2u^2)]+c_3[(\rho+p)\gamma^2\{(1+\gamma^2u^2)(\frac{-\iota\omega}{\alpha}+\iota
ku)+u'(1+\gamma^2u^2)\nonumber\\
&&\times(1+4\gamma^2u^2)+2u\gamma^2(a_z+(1+2\gamma^2u^2)VV')\}
+2\gamma^2u(1+\gamma^2u^2)\nonumber\\
&&\times(\rho'+p')]+c_4[(\rho+p)\gamma^4\{(\frac{-\iota\omega}{\alpha}
+\iota ku)uV+u^2V'(1+4\gamma^2V^2)\nonumber\\
&&+2V(a_z+(1+2\gamma^2u^2)uu')\}+2\gamma^4u^2V(\rho'+p')]=0,\\
\label{49}&&c_1\rho\gamma^2[\frac{-\iota\omega}{\alpha}
+u\{a_z+\gamma^2(VV'+uu')\}]
+c_2p[\frac{-\iota\omega}{\alpha}(\gamma^2-1)\nonumber\\
&&+\gamma^2u\{a_z+\gamma^2(VV'+uu')\}]
+c_3(\rho+p)\gamma^2[\gamma^2u(\frac{-2\iota\omega}{\alpha}+\iota ku)\nonumber\\
&&+(a_z+2\gamma^2uu')(1+2\gamma^2u^2)+\gamma^2VV'(1+4\gamma^2u^2)]
+c_4(\rho+p)\gamma^2\nonumber\\
&&\times[\gamma^2V(\frac{-2\iota\omega}{\alpha}+\iota ku)
+\gamma^2uV'(1+4\gamma^2V^2)+2\gamma^2uV(a_z\nonumber\\
&&+2\gamma^2uu')]=0.
\end{eqnarray}
These equations can be used to find dispersion relations.

\subsection{Numerical Solutions}

For rotating plasma, the velocity assumption can be modified as
follows.
\begin{itemize}
\item  Velocity components: $u=V,~x$ and $z$-components of velocity
are modified as  $u=V=-\sqrt\frac{1-\alpha^2}{2}$.\\\\
Thus, the Lorentz factor becomes
$$\gamma=1/\sqrt{1-u^2-V^2}=1/\alpha.$$
\end{itemize}
This assumption along with the assumptions of time lapse, specific
enthalpy, density and pressure given in Section 4, satisfy the
GRMHD Eqs.(\ref{4})-(\ref{8}) for the region $1.4\leq z\leq10,
0\leq\omega\leq 10$. The determinant of the coefficients of
constants in Eq.(\ref{46})-(\ref{49}) leads to two dispersion
relations. The equation obtained from the real part of the
determinant is quartic, i.e.,
\begin{equation}\label{c}
A_1(z)k^4+A_2(z,\omega)k^3+A_3(z,\omega)k^2+A_4(z,\omega)k+A_5(z,\omega)=0
\end{equation}
which yields four values of $k$ out of which two are real and
other two are complex conjugate of each other. The imaginary part
gives a cubic equation of the form
\begin{equation}\label{d}
B_1(z)k^3+B_2(z,\omega)k^2+B_3(z,\omega)k+B_4(z,\omega)=0
\end{equation}
which leads to three roots in which one is real and other two are
complex conjugate of each other.

The real values of $k$ obtained from Eq.(\ref{c}) and (\ref{d})
along with respective refractive index and its change with respect
to angular frequency are shown in Figures 4, 5 and 6.
\newpage
The results derived from the Figures 4-6 can be represented in the
following tables.
\begin{center}
Table II. Direction and refractive index of waves.
\end{center}
\begin{tabular}{|c|c|c|c|c|}
\hline & \textbf{Direction of waves} & \textbf{Refractive index}
($n$)\\ \hline
& &  $n>1$ and increases in the region\\
\textbf{4} & Move towards the event horizon  & $1.4 \leq z\leq1.8,0\leq\omega\leq 10$\\
& & with the decrease in $z$  \\
\hline
\textbf{5} &  Same as $4$ &  Same as above\\
\hline
& &  $n>1$ and increases in the region\\
\textbf{6} &  Same as $4$ & $1.4 \leq z\leq1.7,0\leq\omega\leq 10$\\
& & with the decrease in $z$  \\
\hline
\end{tabular}
\\
\\
\\
The regions of normal and anomalous dispersion can be summarized
as
\begin{center}
Table III. Regions of dispersion.
\end{center}
\begin{center}
\begin{tabular}{|c|c|c|} \hline
& \textbf{Normal dispersion} & \textbf{Anomalous dispersion}\\
\hline
\textbf{4} &                  ---                     &
$1.4\leq z\leq 10, 0\leq\omega\leq 10$\\
\hline
\textbf{5} & $1.4\leq z\leq 10, 0\leq\omega\leq 10$ &
                  ---                    \\
\hline
\textbf{6} &                  ---                     &
$1.4\leq z\leq 10, 0\leq\omega\leq 10$\\
\hline
\end{tabular}
\end{center}

\section{Rotating Magnetized Hot Plasma}

Here, we assume that the plasma is magnetized and rotating. The
velocity and magnetic field of fluid are assumed to lie in $xz$
plane. The respective Fourier analyzed perturbed GRMHD equations
will remain the same as given by Eqs.(\ref{26})-(\ref{32}) in
Section 3.

\subsection{Numerical Solutions}

We shall use the same values of time lapse, specific enthalpy,
density, pressure, $x$ and $z$-components of velocity as given in
Sections 4 and 5 with the following restrictions on the magnetic
field.
\begin{itemize}
\item When we take $u=V$ and $V^F=0,$ Eq.(\ref{a}) leads to
$\lambda=1$.
\item $B=\sqrt{\frac{176}{7}}$.
\end{itemize}
These restrictions satisfy the GRMHD Eqs.(\ref{4})-(\ref{8}) for the
region $1.4\leq z\leq10, 0\leq\omega\leq 10$. From Eqs.(\ref{27})
and (\ref{28}), we have $c_5=0$. Substituting $c_5=0$ in
Eqs.(\ref{26}) and (\ref{29})-(\ref{32}), we obtain a $5\times5$
matrix of the coefficients of constants. Its determinant leads to
two dispersion relations. The real parts give
\begin{eqnarray}{\setcounter{equation}{1}}
&&A_1(z)k^4+A_2(z,\omega)k^3+A_3(z,\omega)k^2+A_4(z,\omega)k+A_5(z,\omega)=0
\end{eqnarray}
yielding four values of $k$ out of which two are real and
interesting shown in Figures 7 and 8. The imaginary part leads to
\begin{eqnarray}
&&B_1(z)k^5+B_2(z,\omega)k^4+B_3(z,\omega)k^3+B_4(z,\omega)k^2+B_5(z,\omega)k\nonumber\\
&&+B_6(z,\omega)=0.
\end{eqnarray}
This represents fifth order equation giving five roots in which
three are real. These solutions are represented by Figures 9, 10
and 11.\\\\
 Results related to the direction of waves and the
refractive index are given below.
\begin{center}
Table IV. Direction and refractive index of waves.
\end{center}
\begin{tabular}{|c|c|c|c|c|}
\hline & \textbf{Direction of waves} & \textbf{Refractive index}
($n$) \\ \hline
&  & $n>1$ and increases in the region\\
\textbf{ 7}& Moving away from the event horizon &
$1.4\leq z\leq 1.8, 0\leq\omega\leq 10$\\
& & with the decrease in $z$\\
\hline \textbf{ 8}& Same as $7$ &
Same as above\\
\hline
&  & $n>1$ and increases in the region\\
\textbf{ 9 }& Moving towards the event horizon &
$1.4\leq z\leq 1.7, 0\leq\omega\leq 10$\\
& &  with the decrease in $z$\\
\hline \textbf{10}& Same as $9$ &
Same as above\\
\hline \textbf{11}& Same as $9$&
Same as above\\
\hline
\end{tabular}\\\\
\newpage
The information about the regions of normal and anomalous
dispersion obtained from Figures 7-11 is given in the following
table.
\begin{center}
Table V. Regions of dispersion.
\end{center}
\begin{center}
\begin{tabular}{|c|c|c|c|c|}
\hline \textbf{Figure}&  \textbf{Normal dispersion} &
\textbf{Anomalous dispersion} \\ \hline \textbf{7} &
---& $1.4\leq z\leq 10, 0\leq\omega\leq 10$ \\ \hline \textbf{8} &
$3\leq z\leq 10, 1\leq\omega\leq 10$       &                ---
\\\hline & $2\leq z\leq 10, 2\leq\omega\leq 2.4$      &
$1.8\leq z\leq 10, 2.8\leq\omega\leq 3$ \\& $2\leq z\leq
10,3.41\leq\omega\leq 3.6$   & $1.9\leq z\leq 10, 3.6\leq\omega\leq
3.8$  \\& $2\leq z\leq 10, 4.2\leq\omega\leq 4.6$    & $2.3\leq
z\leq 10, 4.68\leq\omega\leq 5$  \\&
$2.3\leq z\leq 10, 5.4\leq\omega\leq 5.62$ &
$1.6\leq z\leq 10, 5.65\leq\omega\leq 6$  \\
\textbf{9} & $2\leq z\leq 10, 6.38\leq\omega\leq 6.6$   &
$1.8\leq z\leq 10, 6.6\leq\omega\leq 6.8$    \\
& $2\leq z\leq 10, 7\leq\omega\leq 7.35$     &
$1.9\leq z\leq 10, 7.38\leq\omega\leq 7.6$\\
& $2\leq z\leq 10, 8\leq\omega\leq 8.2$      &
$2\leq z\leq 10, 8.3\leq\omega\leq 8.97$   \\
& $2\leq z\leq 10, 9\leq\omega\leq 9.2$      &
$1.8\leq z\leq 10, 9.58\leq\omega\leq 9.84$ \\
& $2\leq z\leq 10, 9.9\leq\omega\leq 10$     &
\\\hline
& $2.1\leq z\leq 10, 1.7\leq\omega\leq 2.3$  &
$2\leq z\leq 10, 2.5\leq\omega\leq 2.8$     \\
& $2\leq z\leq 10, 4.21\leq\omega\leq 4.6$   &
$2\leq z\leq 10, 4.61\leq\omega\leq 5$      \\
& $2.3\leq z\leq 10, 5.1\leq\omega\leq 5.4$  &
$2\leq z\leq 10, 5.41\leq\omega\leq 5.65$  \\
& $2\leq z\leq 10, 5.67\leq\omega\leq 6.1$   & $1.8\leq z\leq
10,6.14\leq\omega\leq 6.4$\\
\textbf{10}& $2\leq z\leq 10, 7.21\leq\omega\leq 7.6$   &
$1.8\leq z\leq 10, 6.9\leq\omega\leq 7.2$  \\
& $2\leq z\leq 10, 8.01\leq\omega\leq 8.2$   &
$1.7\leq z\leq 10, 8.6\leq\omega\leq 8.88$  \\
& $2\leq z\leq 10, 8.4\leq\omega\leq 8.6$    &
$1.8\leq z\leq 10, 9.69\leq\omega\leq 9.92$  \\
& $2\leq z\leq 10, 9.41\leq\omega\leq 9.65$  &
\\ \hline
& $1.7\leq z\leq 10, 1.21\leq\omega\leq 1.4$ &
$1.7\leq z\leq 10, 0.4\leq\omega\leq 0.57$  \\
& $2\leq z\leq 10, 2.6\leq\omega\leq 2.77$   &
$2\leq z\leq 10, 3.65\leq\omega\leq 3.85$    \\
& $1.9\leq z\leq 10, 3.4\leq\omega\leq 3.6$  &
$2\leq z\leq 10, 5.7\leq\omega\leq 6.1$     \\
& $2\leq z\leq 10, 4\leq\omega\leq 4.2$      &
$2\leq z\leq 10, 6.44\leq\omega\leq 6.8$   \\
\textbf{11} & $2\leq z\leq 10, 4.61\leq\omega\leq 4.97$  &
$2\leq z\leq 10, 7.3\leq\omega\leq 7.6$    \\
& $2\leq z\leq 10, 5.4\leq\omega\leq 5.69$   &
$2\leq z\leq 10, 7.8\leq\omega\leq 8$       \\
& $2\leq z\leq 10, 6.9\leq\omega\leq 7.2$    &
$1.8\leq z\leq 10, 8.37\leq\omega\leq 8.6$  \\
& $2\leq z\leq 10, 7.6\leq\omega\leq 7.8 $   &
$2.2\leq z\leq 10, 9.5\leq\omega\leq 9.81$  \\
& $2\leq z\leq 10, 8.01\leq\omega\leq 8.35$  &\\
& $2\leq z\leq 10, 9.2\leq\omega\leq 9.4$    &\\
\hline
\end{tabular}
\end{center}
\section{Summary}

In this paper, we find the wave properties of hot plasma in
Schwarzschild magnetosphere. For this purpose, we have derived the
3+1 GRMHD equations for the scenario. Their component and Fourier
analyzed forms are formulated using the specific assumptions.
Dispersion relations are calculated for the non-rotating, rotating
non-magnetized and rotating magnetized plasmas.

For hot plasma living in non-rotating plasma, our assumptions
satisfy the 3+1 GRMHD equations in the region $1.2\leq z \leq10$.
In Figure 1, we have found that the waves are directed away from
the event horizon while Figure 2 shows that the waves move towards
the event horizon. In Figure 3, the waves are directed away from
the event horizon. In all these figures, the dispersion is found
to be normal and anomalous randomly.

For the rotating non-magnetized plasma, the assumed parameters
satisfy the 3+1 GRMHD equations in the region $1.4\leq z\leq10$.
All the figures indicate that the waves are directed towards the
event horizon. In Figure 4, dispersion is anomalous in the whole
region while Figure 5 indicates that dispersion is normal
throughout the region. In Figure 6, dispersion is anomalous.

For the rotating magnetized plasma, our assumptions satisfy the
3+1 GRMHD equations in the region $1.4\leq z\leq 10$. In Figure 7,
the dispersion is anomalous and waves are directed away from the
event horizon. Figure 8 shows that the waves move away from the
event horizon and dispersion is normal in most of the region.
Figures 9, 10 and 11 admit random points of normal dispersion. In
these figures, the waves move towards the event horizon.

The summary of results and comparison with previous literature is
given as follows.

We compare our results with the previous work on isothermal plasma
\cite{32}. We have used here the variable specific enthalpy while in
previous work this is constant. Here propagation vector admits both
positive and negative values which shows that the waves can move
away and towards the event horizon. In the previous work, this
always takes negative values which indicates that the waves move
towards the event horizon. This comparison shows that variation in
specific enthalpy effects the direction of waves. The refractive
index is always greater than one in both cases. Dispersion is normal
and anomalous at random points as shown by figures in both works.
Here the waves move away from the event horizon and disperse
normally in most of the region of figure $8$. This shows that there
is a chance to obtain information and energy from the magnetosphere.
In previous work, waves always move towards the event horizon which
means that no information can be extracted from black hole whether
dispersion is normal or anomalous. The comparison of the results of
cold and hot plasma shows that gas pressure grows the normal
dispersion.
\begin{center}
Table VI. Comparison of results.
\end{center}
\begin{tabular}{|c|c|c|}
\hline
\textbf{Results} & \textbf{Previous work} & \textbf{Recent work} \\
\hline & & \\
\textbf{Direction of waves} & Towards the event horizon & Away from the event \\
& & horizon\\\hline
\textbf{Dispersion} & Normal at random points & Normal in most of the  \\
&& region in Figure 8\\\hline
\textbf{Conclusion} & No information can be & A chance to obtain\\
& extracted & information from \\
& &magnetosphere\\
\hline
\end{tabular}\\\\\\

It is concluded that the hot plasma waves in the Schwarzschild
magnetosphere can have an escape towards the outer end of the
magnetosphere.\\\\
{\bf Acknowledgment:} We would like to thank Dr.
Umber Skeikh for the fruitful discussions during this work.
\renewcommand{\theequation}{A\arabic{equation}}
\section*{Appendix A}

This appendix contains the GRMHD equations for the general line
element and the Schwarzschild planar analogue. The set of Maxwell
equations and conservation laws under the influence of
gravitational field are collectively known as GRMHD equations. The
hot plasma model includes the mass, momentum and energy
conservation laws. The GRMHD equations for the general line
element, given by Eq.(\ref{1}), are \cite{32}
\begin{eqnarray}{\setcounter{equation}{1}}
\label{50}&&\frac{d\textbf{B}}{d\tau}+\frac{1}{\alpha}(\textbf{B}.\nabla)\beta
+\theta\textbf{B}=\frac{1}{\alpha}\nabla\times(\alpha\textbf{V}\times\textbf{B}),\\
\label{51}&&\nabla.\textbf{B}=0,
\end{eqnarray}
\begin{eqnarray}
\label{52}&&\frac{D\rho_0}{D\tau}+\rho_0\gamma^2\textbf{V}.\frac{D\textbf{V}}{D\tau}
+\frac{\rho_0}{\alpha}\left\{\frac{g,_t}{2g}+\nabla.(\alpha\textbf{V}-\beta)\right\}=0,\\
\label{53}&&\left\{\left(\rho_0\mu\gamma^2+\frac{\textbf{B}^2}{4\pi}\right)\gamma_{ij}
+\rho_0\mu\gamma^4V_iV_j
-\frac{1}{4\pi}B_iB_j\right\}\frac{DV^j}{D\tau}\nonumber\\
&&+\rho_0\gamma^2V_i\frac{D\mu}{D\tau}
-\left(\frac{\textbf{B}^2}{4\pi}\gamma_{ij}-\frac{1}{4\pi}B_iB_j\right)
V^j_{|k}V^k=-\rho_0\gamma^2\mu\{a_i\nonumber\\
&&-\frac{1}{\alpha}\beta_{j|i}V^j -(\pounds_t\gamma_{ij})V^j\}
-p_{|i}+\frac{1}{4\pi}(\textbf{V}\times
\textbf{B})_i\nabla.(\textbf{V}\times\textbf{B})\nonumber\\
&&-\frac{1}{8\pi\alpha^2}(\alpha\textbf{B})^2_{|i}+\frac{1}{4\pi\alpha}(\alpha
B_i)_{|j}B^j-\frac{1}{4\pi\alpha}(\textbf{B}\times\{\textbf{V}\times
[\nabla\times\nonumber\\
&&(\alpha\textbf{V}\times\textbf{B})
-(\textbf{B}.\nabla)\beta]+(\textbf{V}\times\textbf{B}).\nabla\beta\})_i,\\
\label{54}&&\frac{d}{d\tau}(\gamma^2\mu\rho_0)-\frac{d
p}{d\tau}+\theta((\gamma^2\rho_0\mu-p)
+\frac{1}{8\pi}((\textbf{V}\times\textbf{B})^2+\textbf{B}^2))\nonumber\\
&&+\frac{1}{2\alpha}\{\gamma^2\rho_0\mu
V^iV^j+p\gamma^{ij}
-\frac{1}{4\pi}((\textbf{V}\times\textbf{B})^i(\textbf{V}\times\textbf{B})^j\nonumber\\
&&+B^iB^j)+\frac{1}{8\pi}((\textbf{V}\times\textbf{B})^2
+\textbf{B}^2)\gamma^{ij}\}\pounds_t\gamma_{ij}+
\mu\rho_0\gamma^2\textbf{V}.\textbf{a}\nonumber\\
&&+\mu\rho_0\gamma^4\textbf{V}.(\textbf{V}.\nabla)\textbf{V}-
\frac{1}{2\pi}\textbf{a}.((\textbf{V}\times\textbf{B})\times\textbf{B})
-\frac{1}{4\pi}(\nabla\times\nonumber\\
&&(\textbf{V}\times\textbf{B})).\textbf{B}
-\frac{1}{\alpha}\left\{\mu\rho_0\gamma^2\textbf{V}.(\textbf{V}.\nabla)\beta+p(\nabla.\beta)
-\frac{1}{4\pi}\textbf{B}.(\textbf{B}.\nabla)\beta
\right.\nonumber\\
&&-\frac{1}{4\pi}(\textbf{V}\times\textbf{B}).\{(\textbf{V}\times\textbf{B}).\nabla\}\beta\left.
+\frac{1}{8\pi}((\textbf{V}\times\textbf{B})^2
+\textbf{B}^2)(\nabla.\beta)\right\}\nonumber\\
&&+\frac{1}{4\pi}\left[(\textbf{V}\times\textbf{B}).(\textbf{a}\times\textbf{B})
+2(\textbf{V}\times\textbf{B}).\frac{d}{d\tau}(\textbf{V}\times\textbf{B})\right.\nonumber\\
&&-\frac{1}{\alpha}(\textbf{V}\times\textbf{B})
.\{(\textbf{V}\times\textbf{B}).\nabla\}\beta
+\theta(\textbf{V}\times\textbf{B}).(\textbf{V}\times\textbf{B})\nonumber\\
&&\left.+\textbf{B}.\frac{d\textbf{B}}{d\tau}\right]=0.
\end{eqnarray}

For the Schwarzschild planar analogue, $\beta$, $\theta$ and
$\pounds_t\gamma_{ij}$ vanish, the perfect GRMHD equations take the
following form
\begin{eqnarray}\label{54}
&&\frac{\partial\textbf{B}}{\partial t}=\nabla \times(\alpha
\textbf{V}\times\textbf{B}),\\
\label{55} &&\nabla.\textbf{B}=0,
\end{eqnarray}
\begin{eqnarray}
\label{56} &&\frac{\partial\rho_0}{\partial
t}+(\alpha\textbf{V}.\nabla)\rho_0+\rho_0\gamma^2
\textbf{V}.\frac{\partial\textbf{V}}{\partial
t}+\rho_0\gamma^2\textbf{V}.(\alpha\textbf{V}.\nabla)\textbf{V}\nonumber\\
&&+\rho_0{\nabla.(\alpha\textbf{V})}=0,\\
\label{57}&&\{(\rho_0\mu\gamma^2+\frac{\textbf{B}^2}{4\pi})\delta_{ij}
+\rho_0\mu\gamma^4V_iV_j
-\frac{1}{4\pi}B_iB_j\}(\frac{1}{\alpha}\frac{\partial}{\partial
t}+\textbf{V}.\nabla)V^j\nonumber\\
&&-(\frac{\textbf{B}^2}{4\pi}\delta_{ij}-\frac{1}{4\pi}B_iB_j)
V^j,_kV^k+\rho_0\gamma^2V_i\{\frac{1}{\alpha}\frac{\partial
\mu}{\partial
t}+(\textbf{V}.\nabla)\mu\}\nonumber\\
&&=-\rho_0\mu\gamma^2a_i-p,_i+
\frac{1}{4\pi}(\textbf{V}\times\textbf{B})_i\nabla.(\textbf{V}\times\textbf{B})
-\frac{1}{8\pi\alpha^2}(\alpha\textbf{B})^2,_i\nonumber\\
&&+\frac{1}{4\pi\alpha}(\alpha B_i),_jB^j-\frac{1}{4\pi\alpha}
[\textbf{B}\times\{\textbf{V}\times(\nabla\times(\alpha\textbf{V}
\times\textbf{B}))\}]_i,\\
\label{58} &&\gamma^2\frac{1}{\alpha}\frac{\partial}{\partial
t}(\mu\rho_0)-\frac{1}{\alpha}\frac{\partial p}{\partial t}
+\rho_0\mu\gamma^2(\textbf{V}.\textbf{a})
+\rho_0\mu\gamma^4\textbf{V}.(\textbf{V}.\nabla)\textbf{V}\nonumber\\
&&-\frac{1}{2\pi}\textbf{a}.\{(\textbf{V}\times\textbf{B})\times\textbf{B}\}
-\frac{1}{4\pi}(\nabla\times(\textbf{V}\times\textbf{B})).\textbf{B}\nonumber\\
&&+\frac{1}{4\pi}\left\{2\frac{1}{\alpha}(\textbf{V}\times\textbf{B}).\frac
{\partial}{\partial t}(\textbf{V}\times\textbf{B})+
(\textbf{V}\times\textbf{B}).(\textbf{a}\times\textbf{B})\right.\nonumber\\
&&\left.+\textbf{B}.\frac{1}{\alpha}\frac {\partial
\textbf{B}}{\partial t}\right\}=0.
\end{eqnarray}

\newpage
\begin{figure}
\begin{tabular}{cccc}
& A & B & C \\
& \epsfig{file=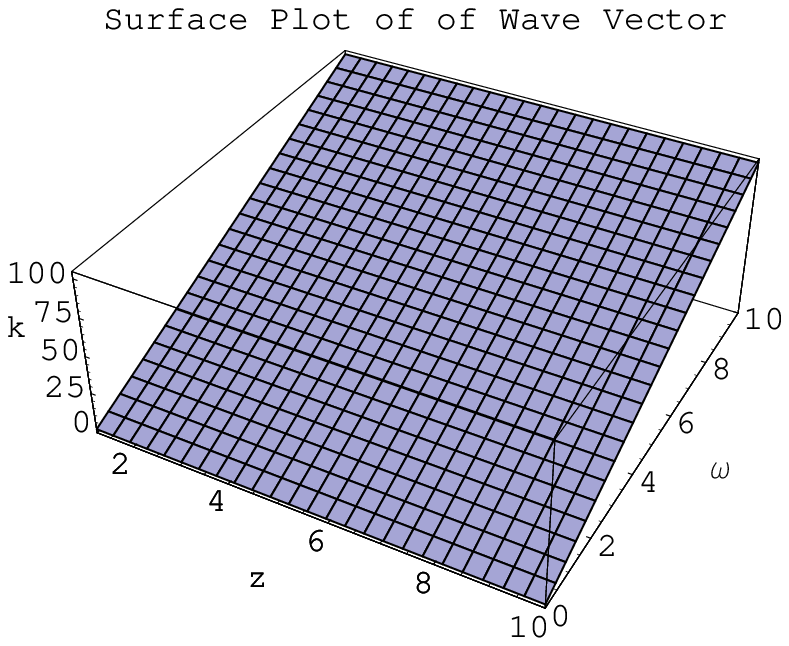,width=0.30\linewidth} &
\epsfig{file=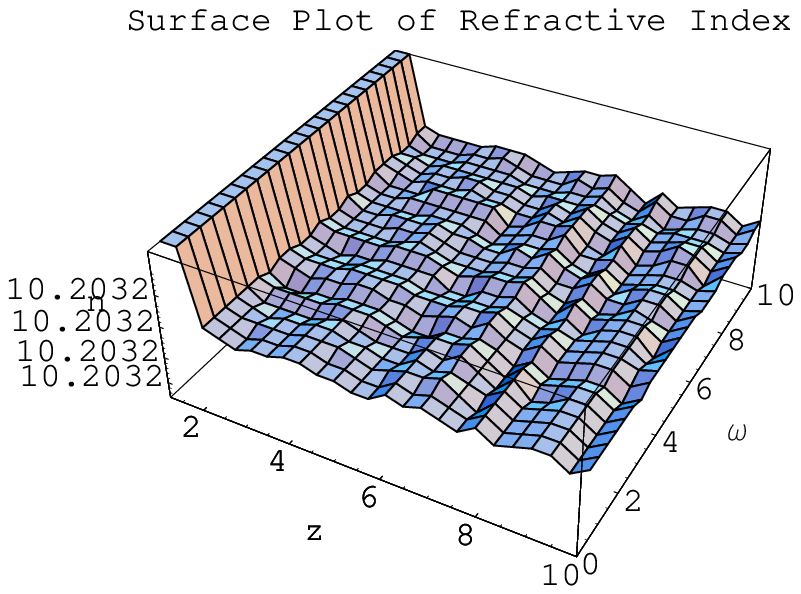,width=0.30\linewidth} &
\epsfig{file=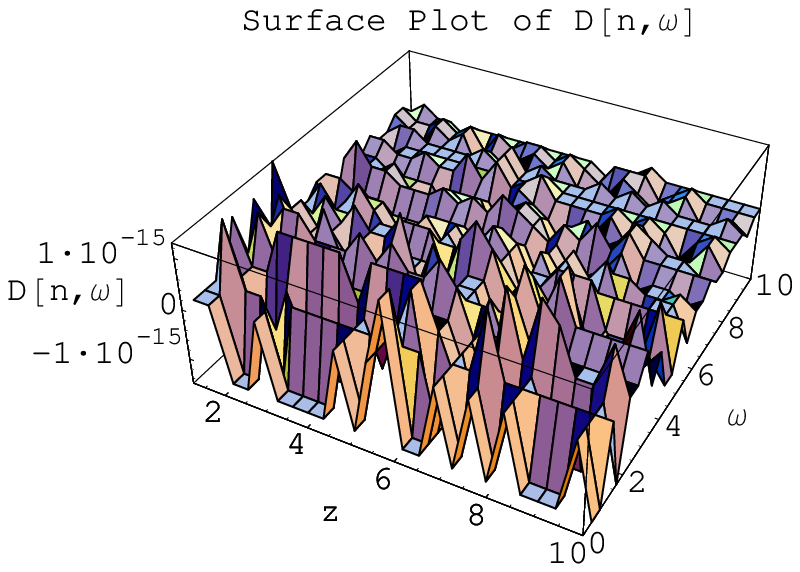,width=0.30\linewidth}\\
\end{tabular}\\\\
\caption{The waves are directed away from the event horizon. The
dispersion is found to be normal and anomalous randomly}
\end{figure}
\begin{figure}
\begin{tabular}{cccc}
& A & B & C \\
& \epsfig{file=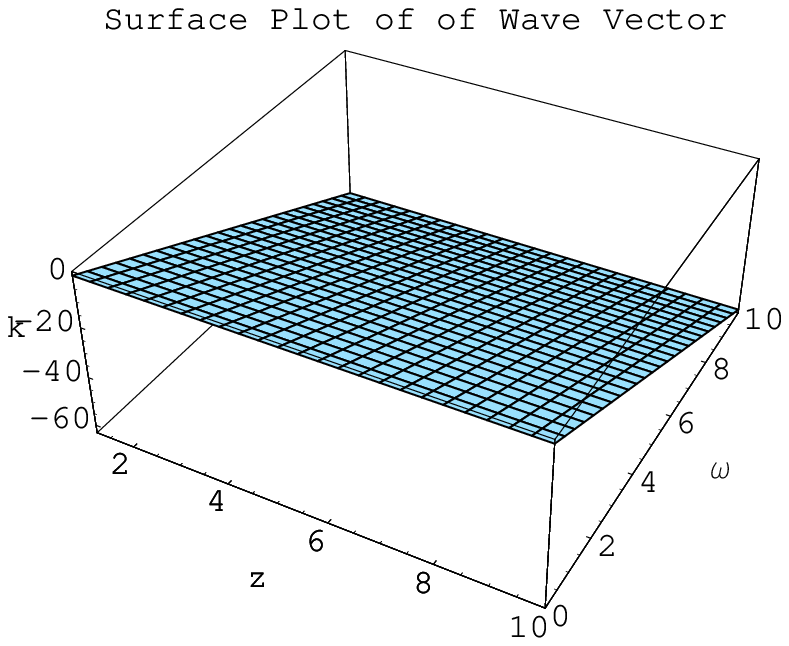,width=0.30\linewidth} &
\epsfig{file=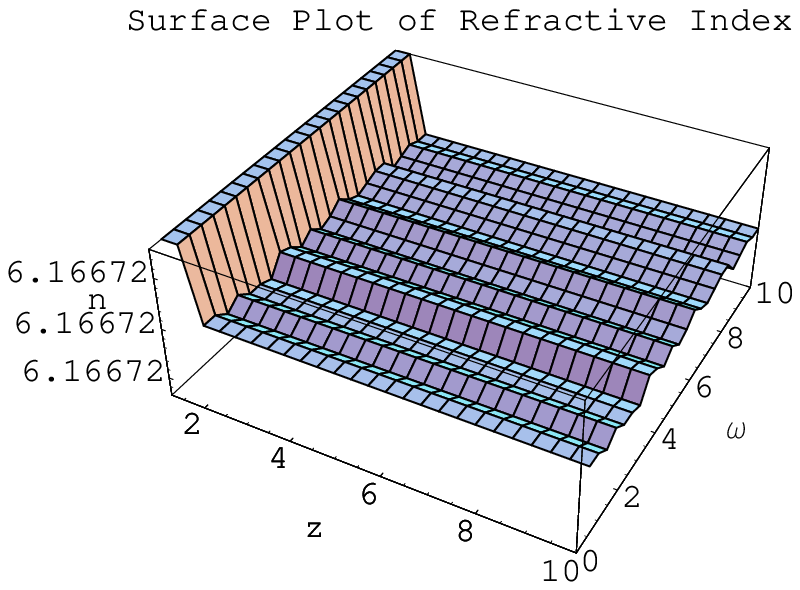,width=0.30\linewidth}&
\epsfig{file=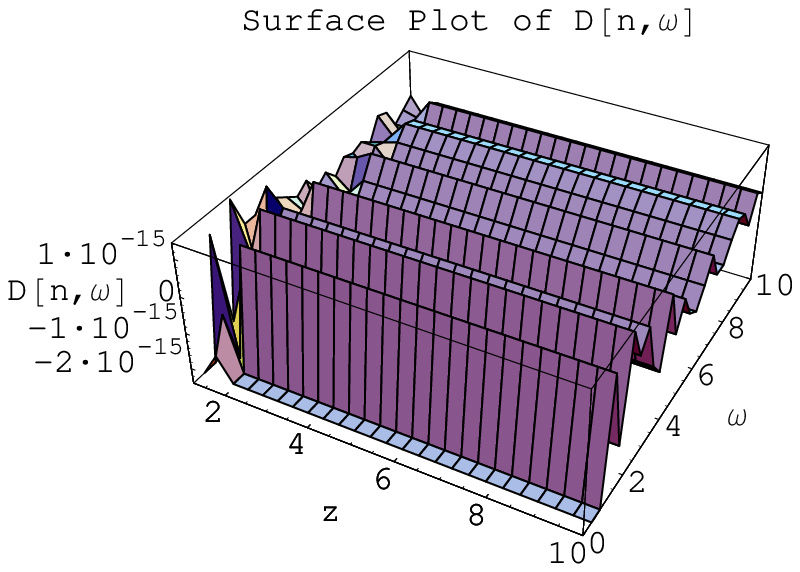,width =0.30\linewidth} \\
\end{tabular}\\\\
\caption{The waves moves towards the event horizon. The dispersion
is normal at random points}
\end{figure}
\begin{figure}
\begin{tabular}{cccc}
& A & B & C \\
& \epsfig{file=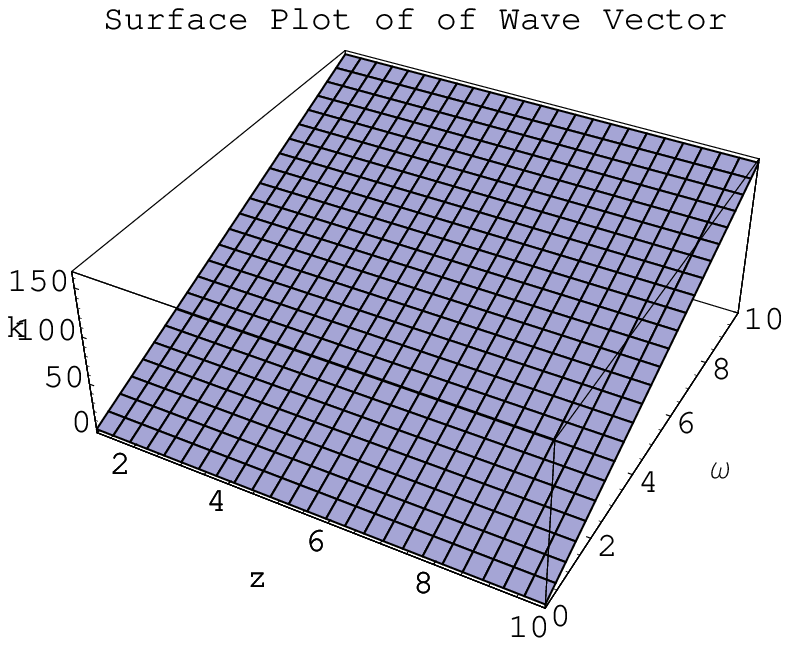,width=0.30\linewidth} &
\epsfig{file=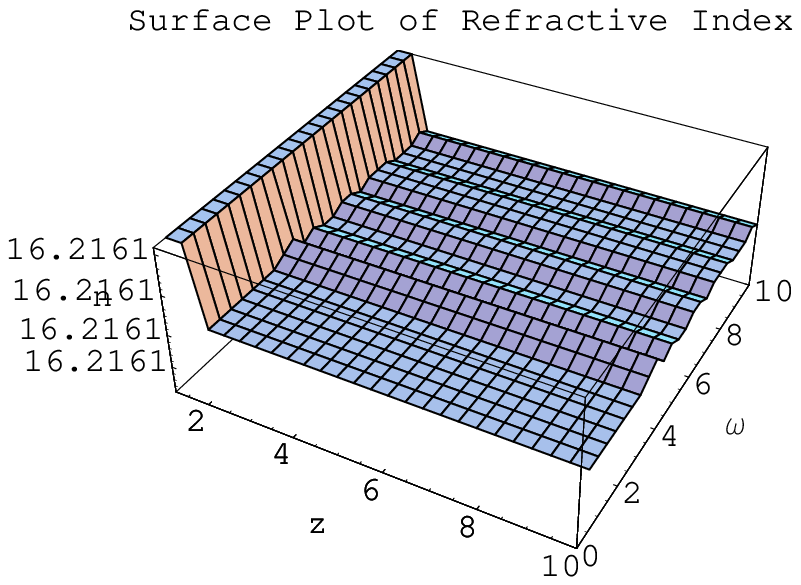,width=0.30\linewidth} &
\epsfig{file=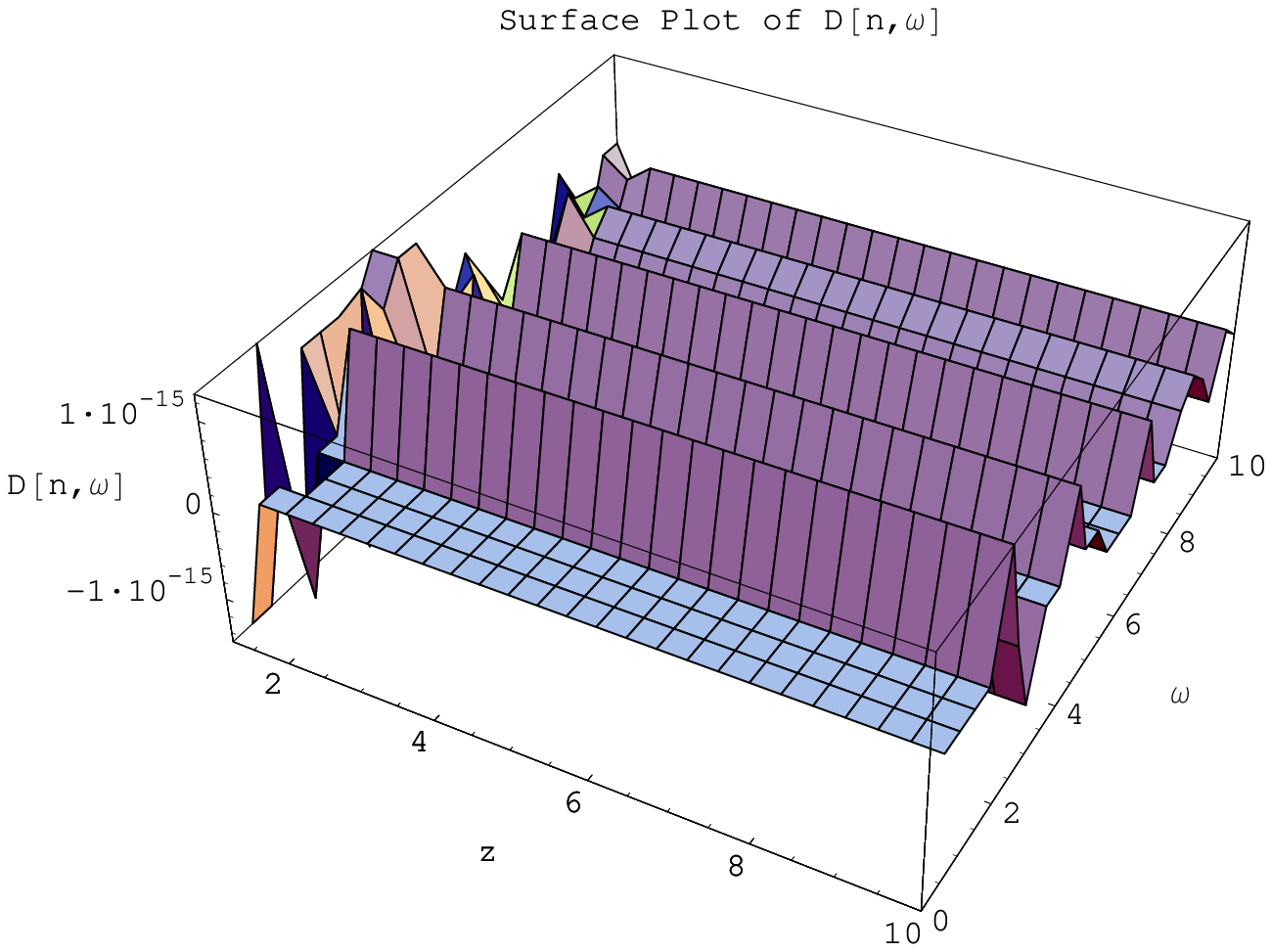,width=0.30\linewidth} \\
\end{tabular}\\\\
\caption{The waves move away from the event horizon. The
dispersion is found to be normal and anomalous at random points}
\end{figure}
\begin{figure}
\begin{tabular}{cccc}
& A & B & C \\
 & \epsfig{file=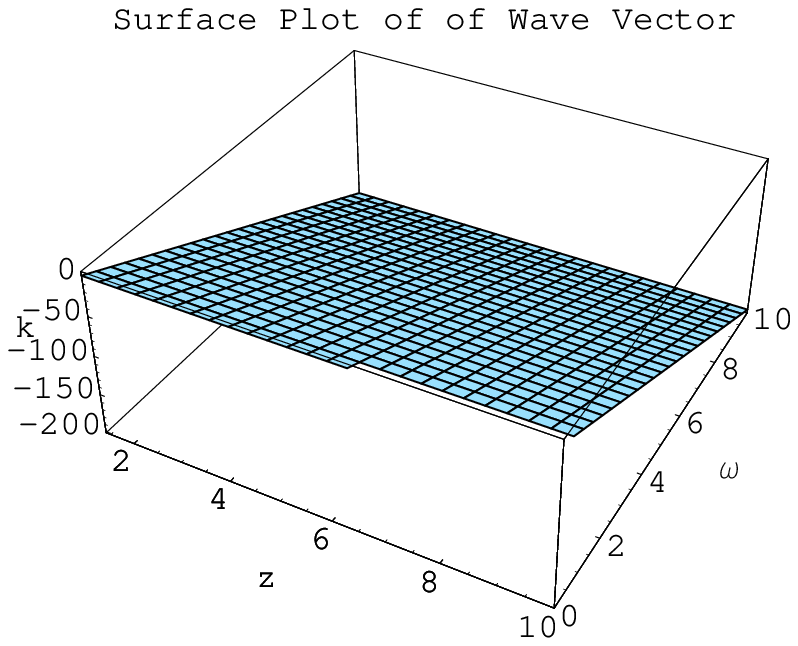,width=0.3\linewidth} &
\epsfig{file=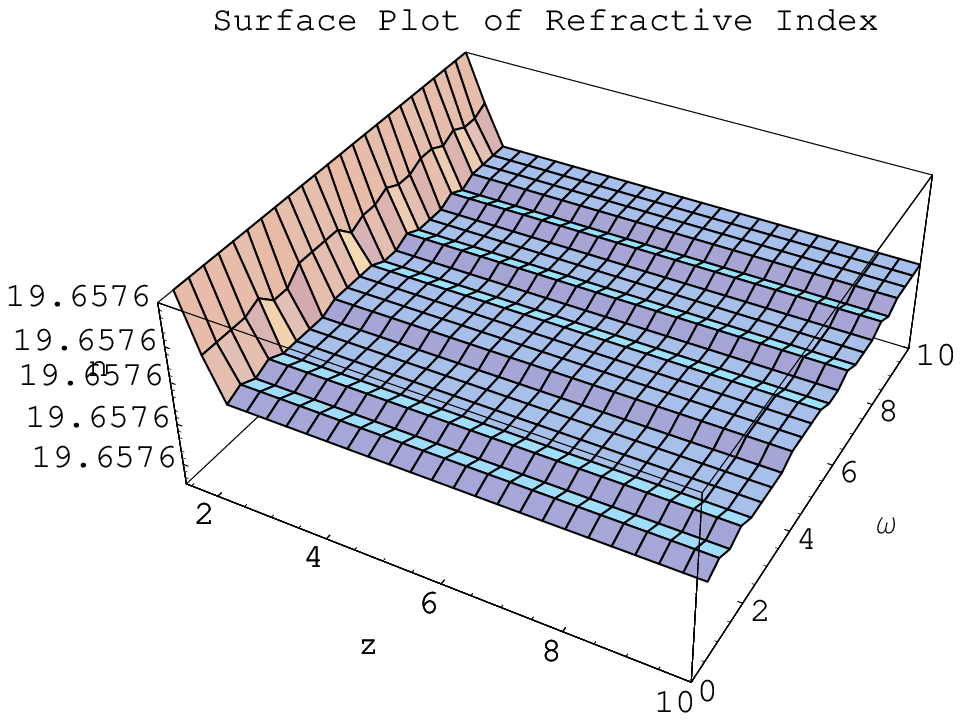,width=0.3\linewidth} &
\epsfig{file=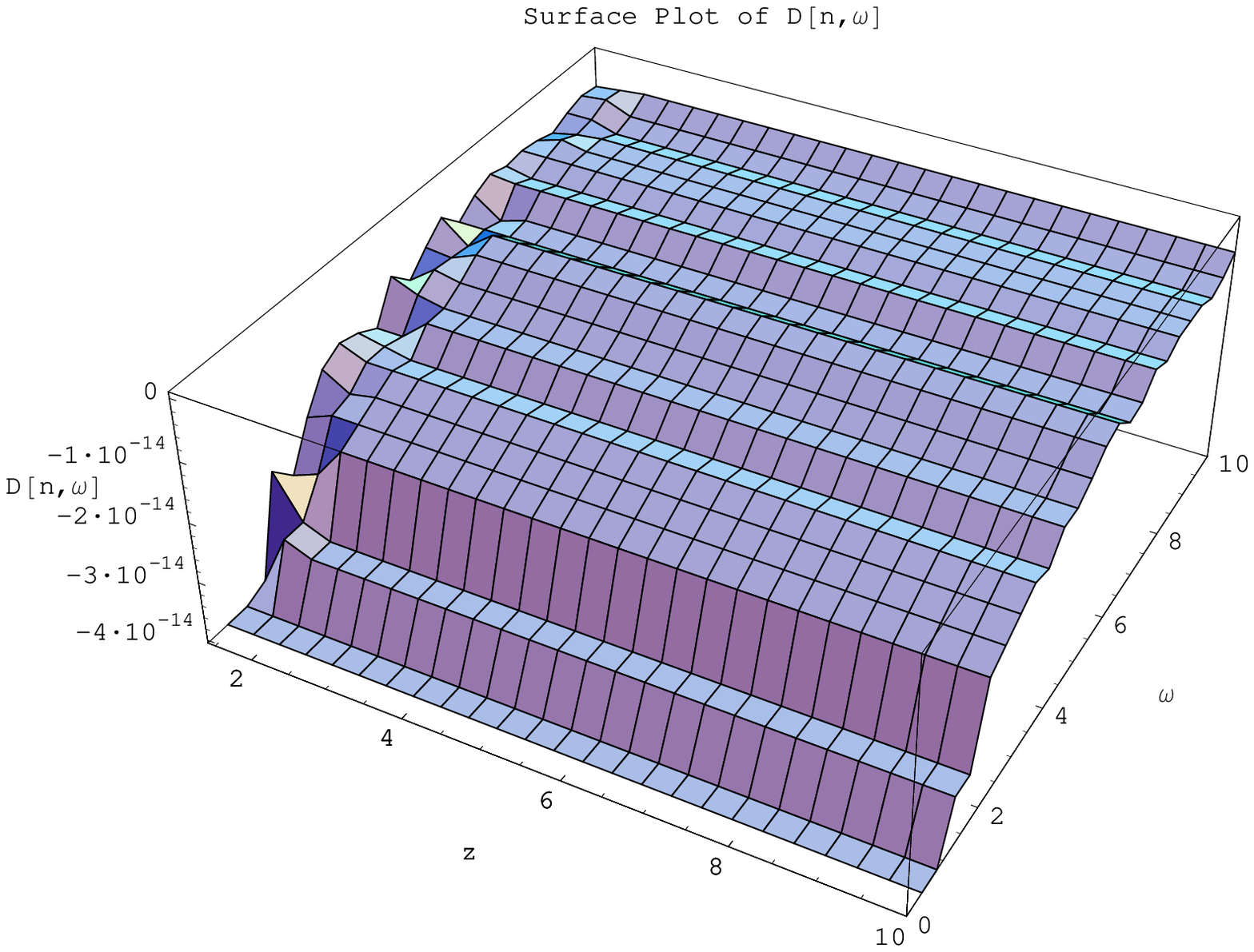,width=0.3\linewidth}\\
\end{tabular}\\\\
\caption{The waves are directed towards the the event horizon. The
region shows anomalous dispersion}
\end{figure}
\begin{figure}
\begin{tabular}{cccc}
& \epsfig{file=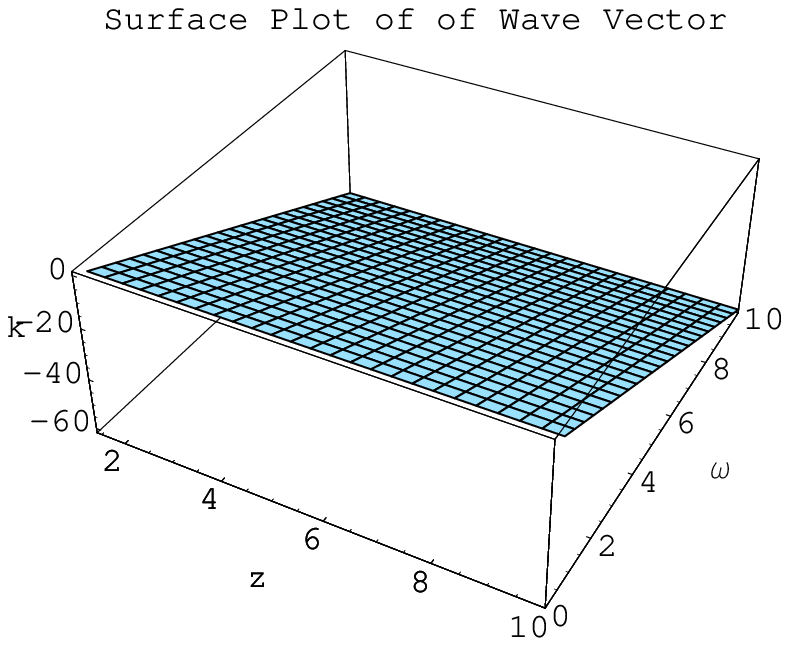,width=0.3\linewidth} &
\epsfig{file=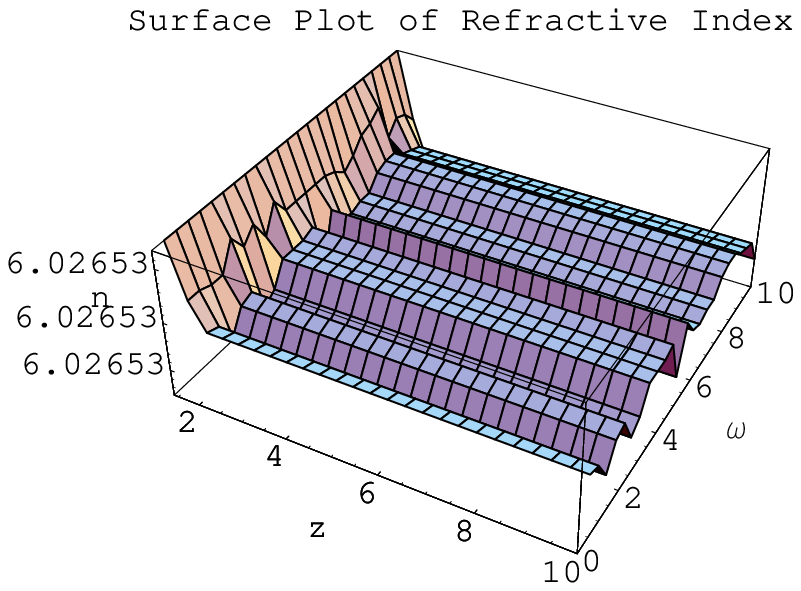,width=0.3\linewidth}&
\epsfig{file=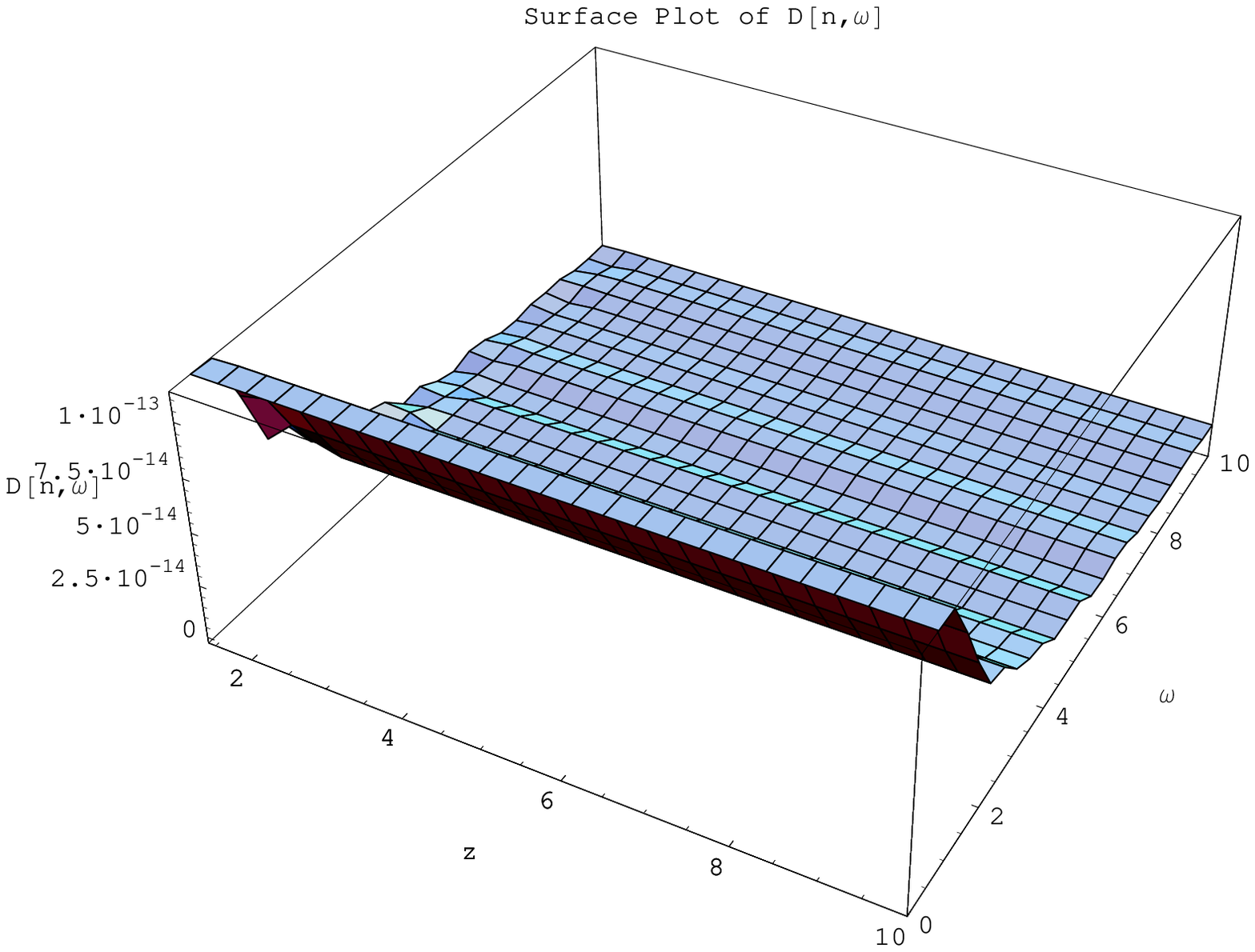,width=0.3\linewidth} \\
\end{tabular}\\\\
\caption{The waves are directed towards the the event horizon. The
dispersion is normal}
\end{figure}
\begin{figure}
\begin{tabular}{cccc}
& \epsfig{file=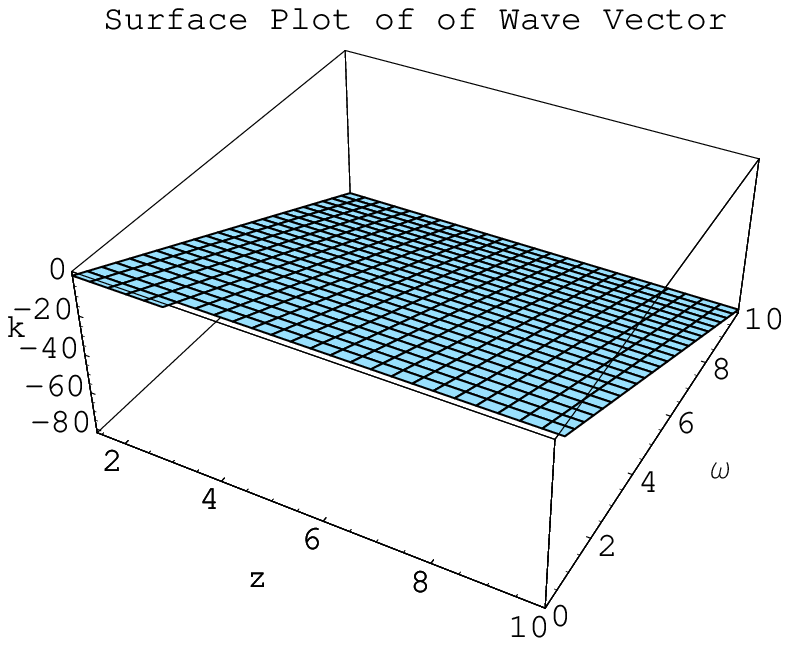,width=0.3\linewidth} &
\epsfig{file=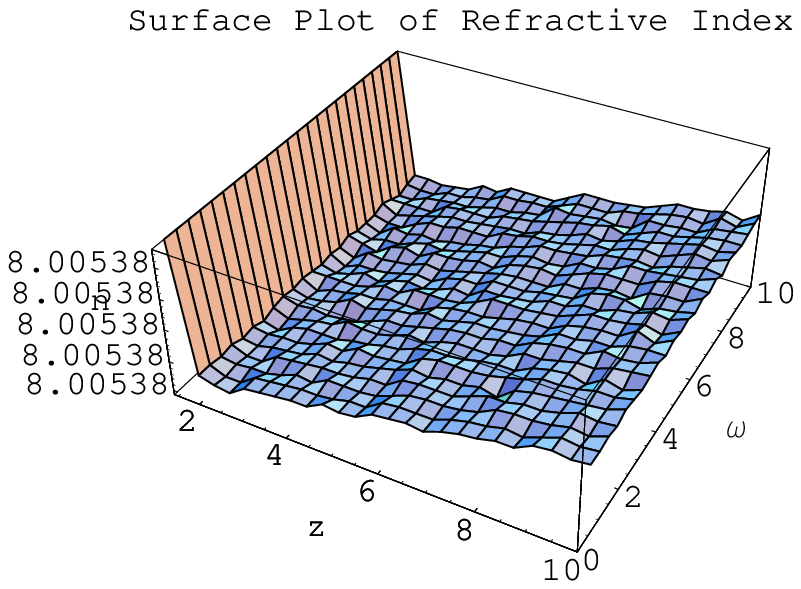,width=0.3\linewidth} &
\epsfig{file=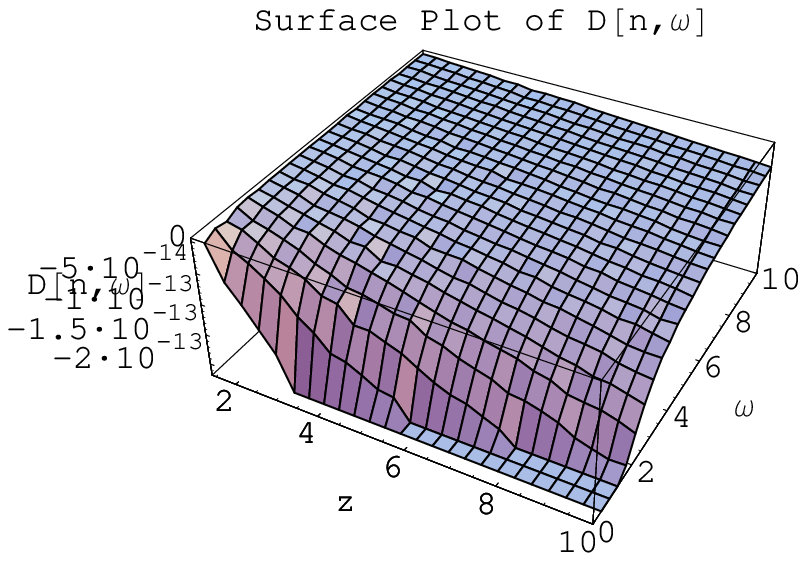,width=0.3\linewidth} \\
\end{tabular}\\\\
\caption{The waves are directed towards the the event horizon. The
whole region admits anomalous dispersion}
\end{figure}
\begin{figure}
\begin{tabular}{cccc}
& A & B & C \\
& \epsfig{file=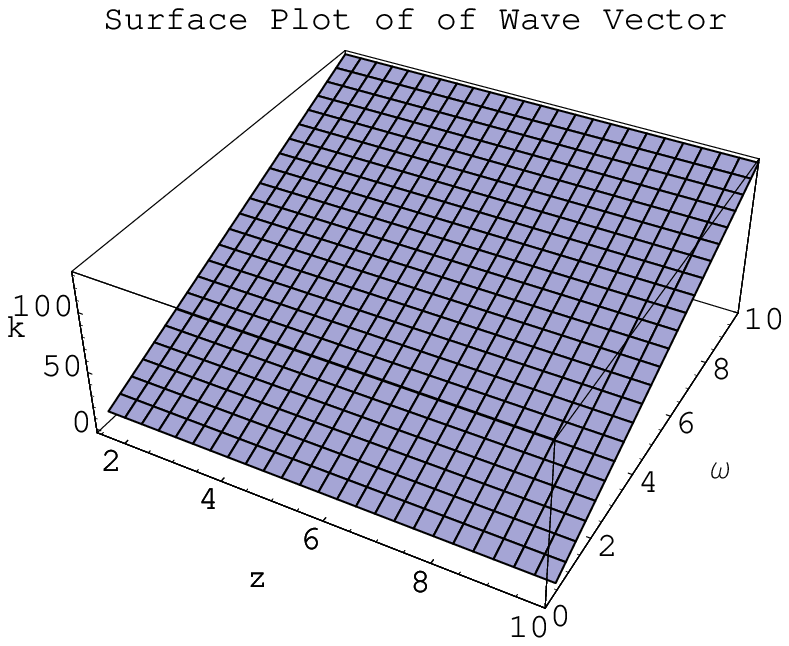,width=0.27\linewidth} &
\epsfig{file=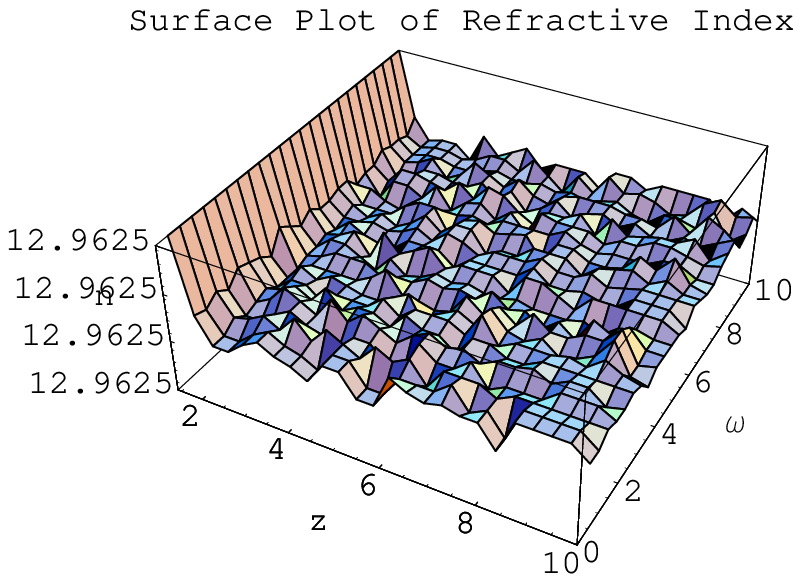,width=0.27\linewidth} &
\epsfig{file=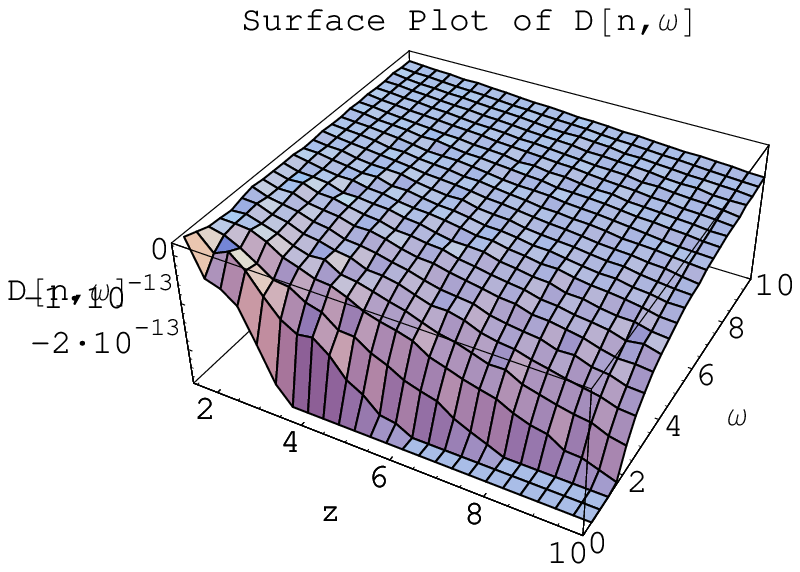,width=0.27\linewidth}\\
\end{tabular}\\\\
\caption{The waves move away from the event horizon. The whole
region admits the anomalous dispersion}
\end{figure}
\begin{figure}
\begin{tabular}{cccc}
& A & B & C \\
& \epsfig{file=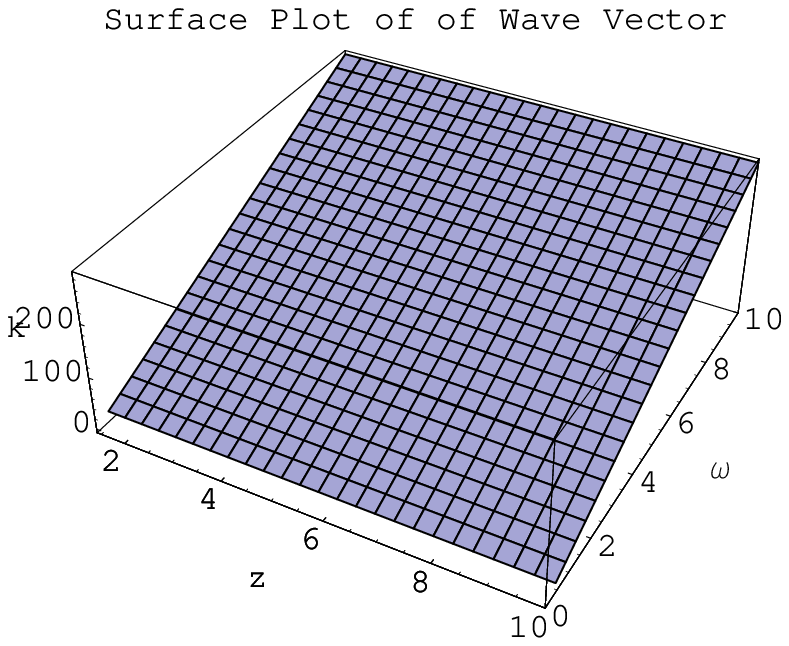,width=0.3\linewidth} &
\epsfig{file=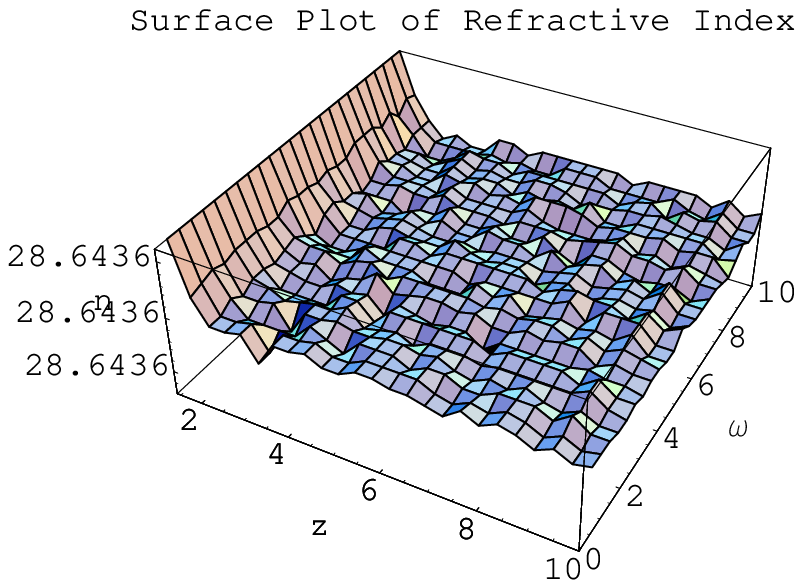,width=0.3\linewidth}&
\epsfig{file=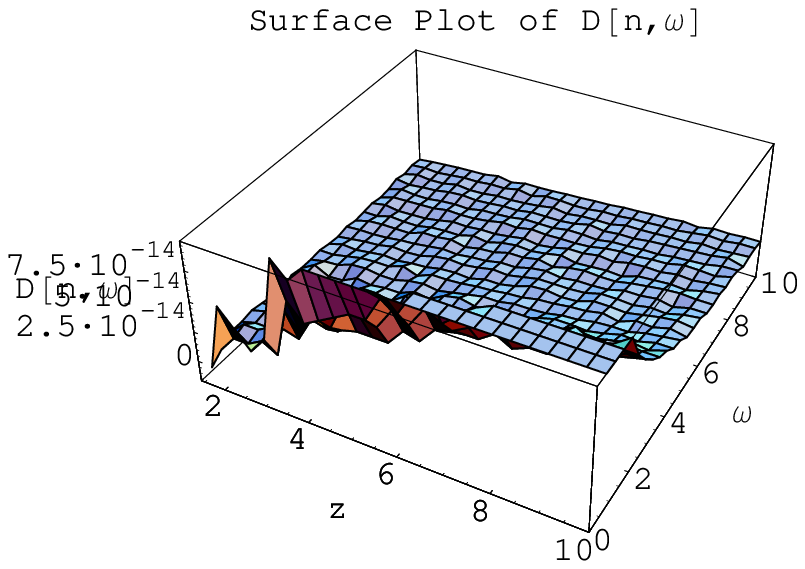,width=0.3\linewidth} \\
\end{tabular}\\\\
\caption{The waves are directed away from the event horizon. The
dispersion is normal in most of the region}
\end{figure}
\begin{figure}
\begin{tabular}{cccc}
& A & B & C \\
& \epsfig{file=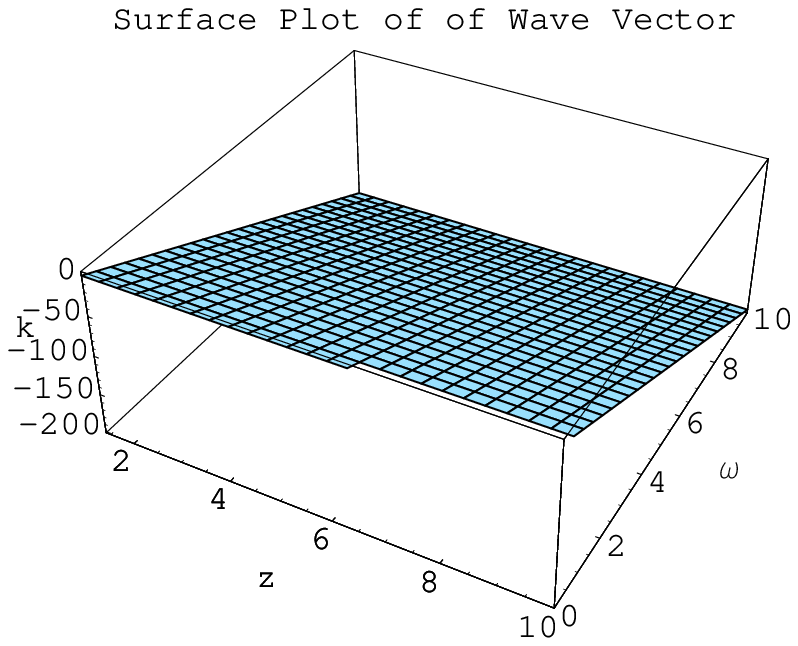,width=0.3\linewidth} &
\epsfig{file=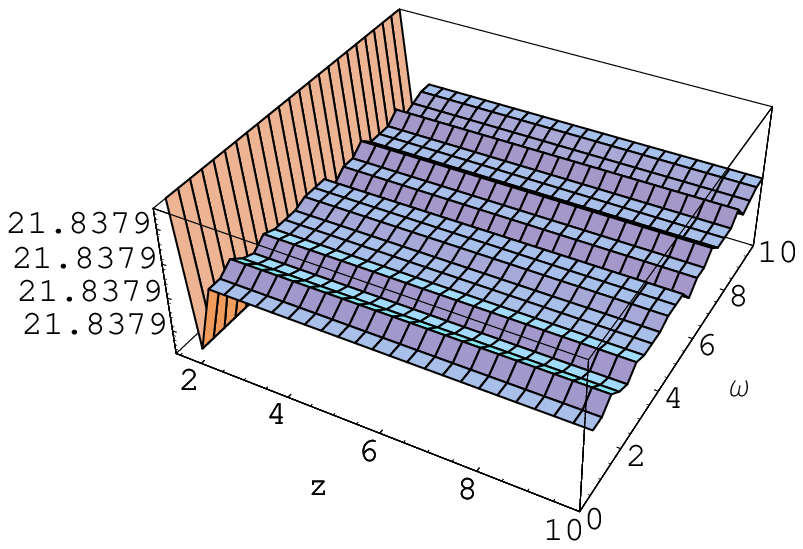,width=0.3\linewidth} &
\epsfig{file=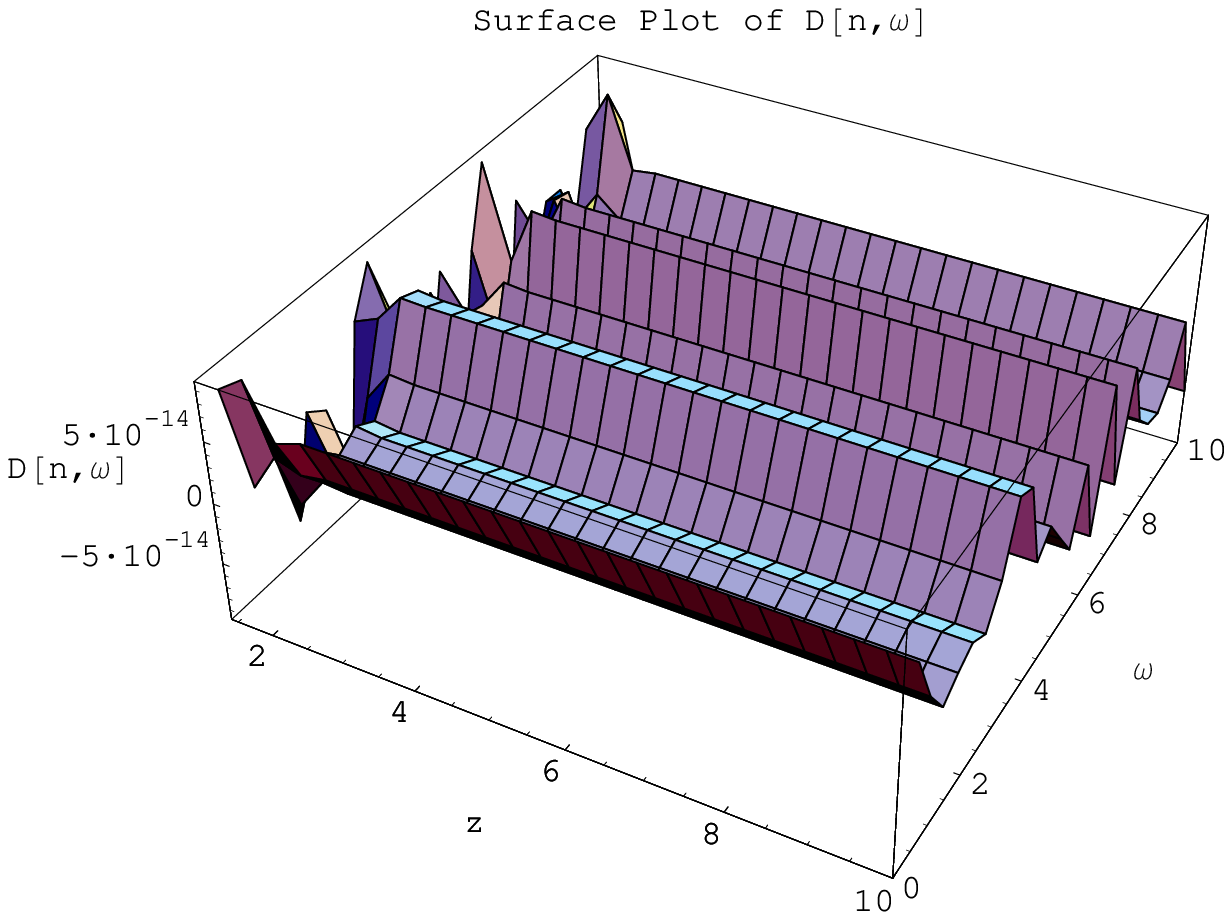,width=0.3\linewidth} \\
\end{tabular}\\\\
\caption{The waves are directed towards the event horizon. Normal
dispersion is found at random points}
\end{figure}
\begin{figure}
\begin{tabular}{cccc}
& A & B & C \\
& \epsfig{file=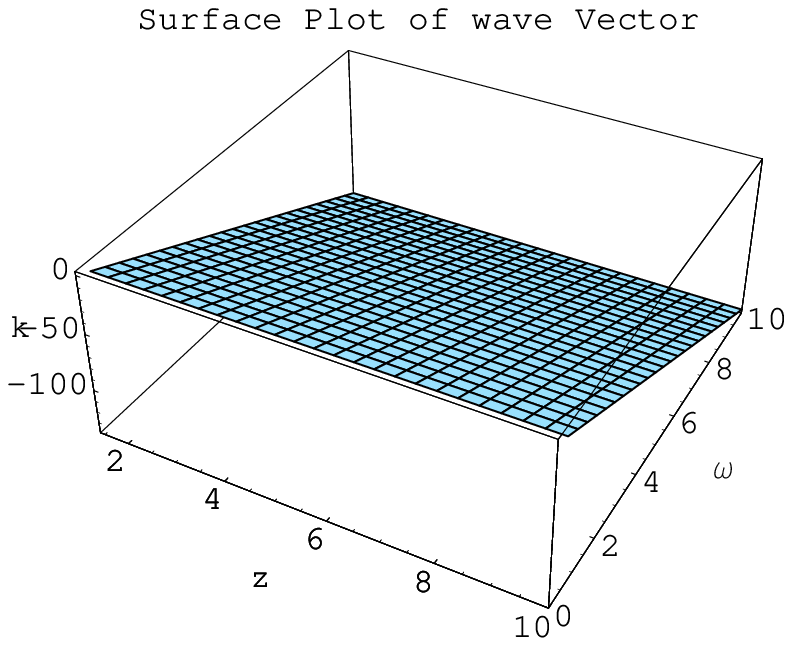,width=0.3\linewidth} &
\epsfig{file=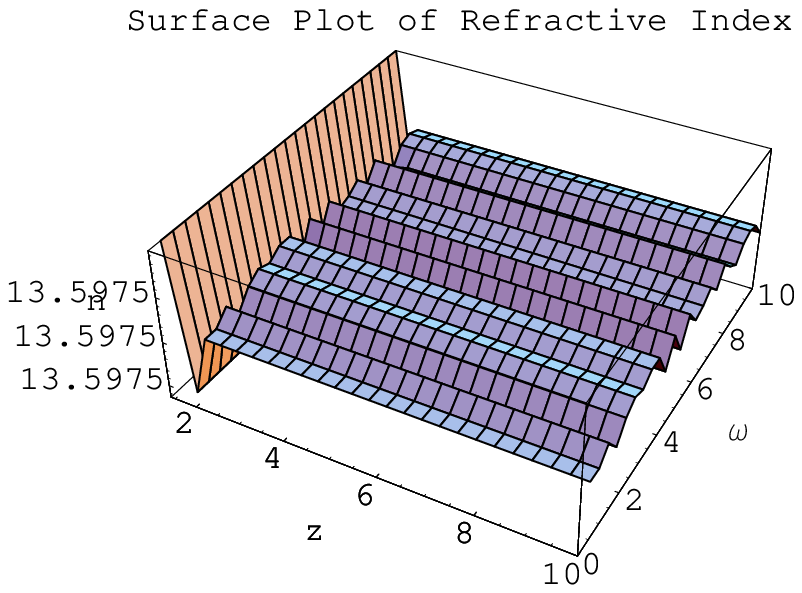,width=0.3\linewidth} &
\epsfig{file=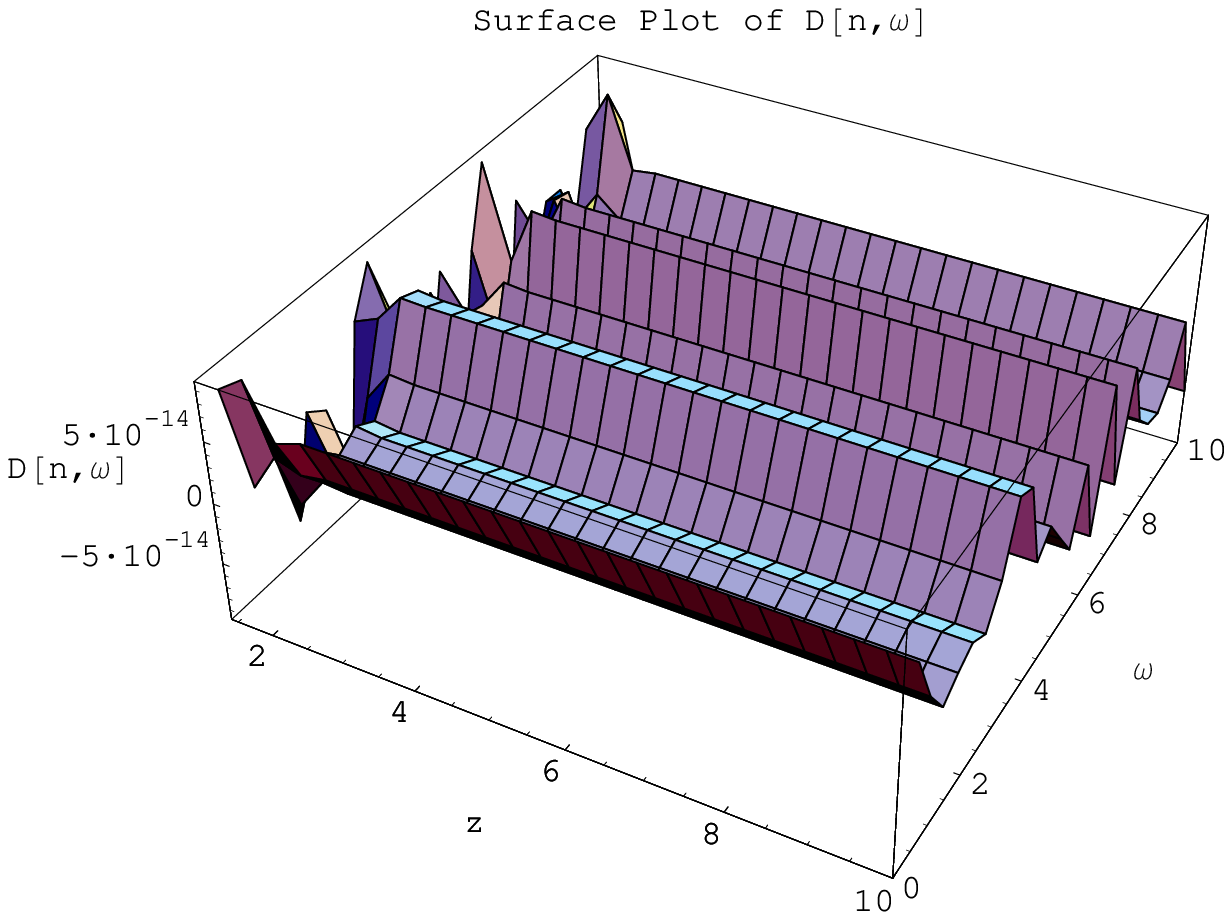,width=0.3\linewidth}\\
\end{tabular}\\
\caption{The waves move towards the event horizon. The whole
region admits the random points of normal and anomalous
dispersion}
\end{figure}
\begin{figure}
\begin{tabular}{cccc}
& A & B & C \\
& \epsfig{file=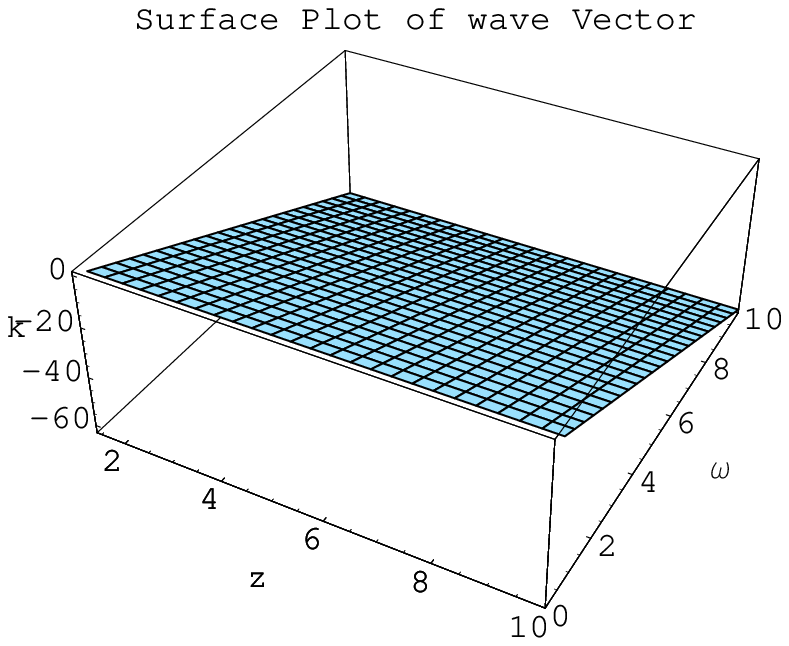,width=0.32\linewidth} &
\epsfig{file=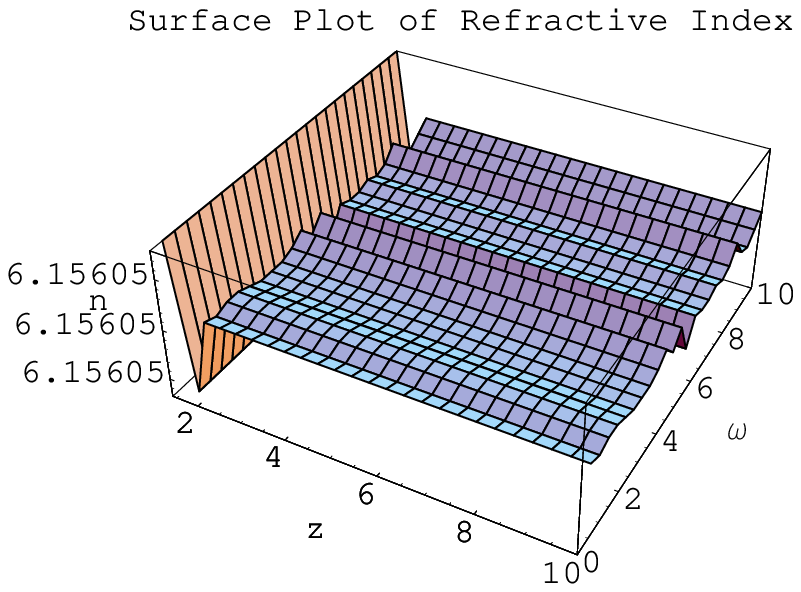,width=0.27\linewidth}&
\epsfig{file=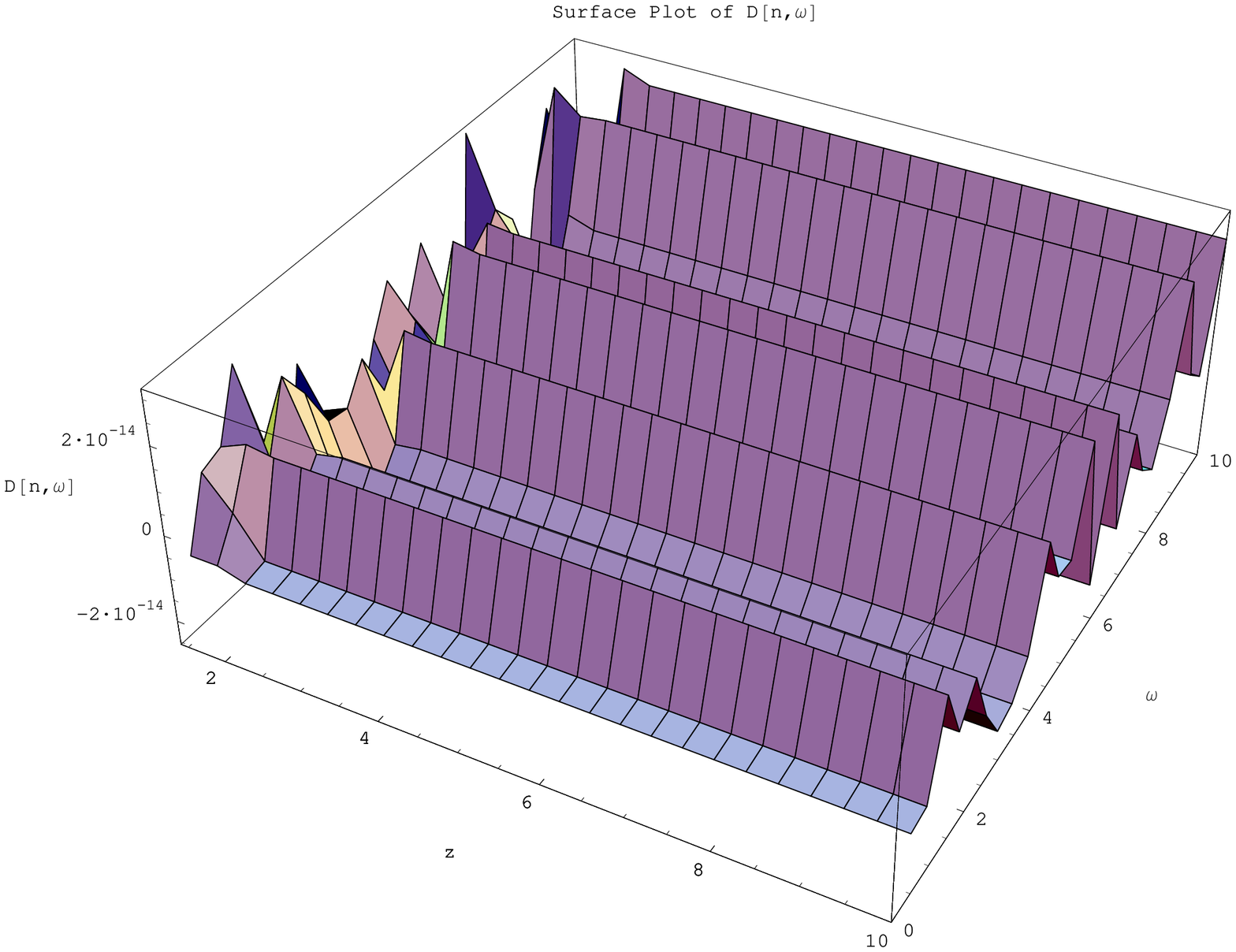,width=0.27\linewidth} \\
\end{tabular}
\caption{The waves are directed towards the the event horizon. The
dispersion is normal at random points}
\end{figure}

\end{document}